\documentclass[aps,prd,twocolumn,reprint,preprintnumbers,superscriptaddress]{revtex4-1}
\usepackage{graphics}  
\usepackage{epsfig}
\usepackage[usenames,dvipsnames]{xcolor}
\usepackage{verbatim}
\usepackage{amsmath}
\usepackage{slashed} 
\definecolor{linkcolor}{RGB}{55,57,154}
\usepackage[pdfstartview={XYZ},colorlinks=true,allcolors=linkcolor]{hyperref} 
\begin{document}

\title{Leptogenesis via Higgs Relaxation}

\author{Louis Yang}
\affiliation{Department of Physics and Astronomy, University of California, Los Angeles, CA 90095-1547, USA}

\author{Lauren Pearce}
\affiliation{William I. Fine Theoretical Physics Institute, School of Physics and Astronomy, University of Minnesota, Minneapolis, MN 55455 USA}

\author{Alexander Kusenko}
\affiliation{Department of Physics and Astronomy, University of California, Los Angeles, CA 90095-1547, USA}
\affiliation{Kavli Institute for the Physics and Mathematics of the Universe (WPI), University of Tokyo, Kashiwa, Chiba 277-8568, Japan}

\preprint{FTPI-MINN-15/27}

\begin{abstract}
An epoch of Higgs relaxation may occur in the early universe during or immediately following postinflationary reheating.  It has recently been pointed out that leptogenesis may occur in minimal extensions of the Standard Model during this epoch~\cite{Kusenko:2014lra}.  We analyse Higgs relaxation taking into account the effects of perturbative and non-perturbative decays of the Higgs condensate, and we present a detailed derivation of the relevant kinetic equations and of the relevant particle interaction cross sections.  We identify the parameter space in which a sufficiently large asymmetry is generated.
\end{abstract}

\maketitle

\section{Introduction}

During the inflationary era, the Higgs field may develop a stochastic distribution of vacuum expectation values (VEVs) due to the flatness of its potential~\cite{Bunch:1978yq,Hawking:1981fz,Linde:1982uu,Starobinsky:1982ee,
Vilenkin:1982wt,Starobinsky:1994bd,Enqvist:2013kaa,Enqvist:2014bua}, or it may be trapped in a quasi-stable minimum.  In both cases, the Higgs field relaxes to its vacuum state via a coherent motion, during which time the Sakharov conditions~\cite{Sakharov:1967dj}, necessary for baryogenesis, are satisfied by the time-dependent Higgs condensate and the lepton-number-violating Majorana masses in the neutrino sector.  At large VEVs, the Higgs field may be sensitive to physics beyond the Standard Model, which can generate an effective chemical potential which increases the energy of antileptons in comparison to leptons.  In Ref.~\cite{Kusenko:2014lra}, we used a $\mathcal{O}_6$ operator familiar from spontaneous baryogenesis models (e.g., \cite{Dine:1990fj}) to produce a baryon asymmetry matching cosmological observations.

In this work, we build on our previous analysis.  In particular, we replace the estimate of the Higgs-neutrino cross section with a tree-level calculation which includes resonant effects.  Additionally, we include the effects of Higgs condensate decay, with both perturbative and non-perturbative contributions.  We also present a detailed derivation of the relevant Boltzmann equation.  Additionally, the effective $\mathcal{O}_6$ operator can be generated through fermionic loops; therefore, its scale can be set either by some heavy mass scale or by the temperature of the plasma.  We consider additional combinations of these scales along with mechanisms to generate the large Higgs VEV during inflation, and in particular, we also present an analysis of the relevant parameter space.

We note that as in Ref.~\cite{Kusenko:2014lra}, we consider an asymmetry produced via the scattering of neutrinos and Higgs bosons in the plasma produced by the decays of the inflaton, and therefore this scenario requires a relatively fast reheating.  This is in contrast to Ref.~\cite{Pearce:2015nga}, which similarly considered the same $\mathcal{O}_6$ operator but produced the matter asymmetry via the decay of the Higgs condensate.

While we focus here on the relaxation of the Higgs field, it has also been observed that the axion field can undergo a similar post-inflationary relaxation~\cite{Chiba:2003vp,Kusenko:2014uta}, and our analysis can easily be extended to this scenario.

The structure of this paper is as follows. In the next section, we consider two specific mechanisms by which the Higgs field can acquire a large vacuum expectation value during inflation.    In the scenarios considered here, the subsequent evolution of the Higgs VEV produces an effective chemical potential, which influences the interactions of leptons in the thermal plasma produced via reheating.  The presence of the plasma, however, also influences the evolution of the Higgs VEV through finite temperature corrections to the effective potential.  Therefore, we discuss reheating in section \ref{sec:Reheat} before we consider the evolution of the Higgs condensate in section \ref{sec:Relaxation}.
Next, we introduce a higher-dimensional operator, involving only Standard Model fields, which represents new physics at some high energy scale. In section \ref{sec:chemical_potential}, we demonstrate that, while the Higgs VEV is in motion, this operator induces an effective chemical potential which distinguishes leptons from antileptons.  We derive the resulting Boltzmann equation for lepton number in section \ref{sec:boltzmann_equation_ahhhh!}.  In section \ref{sec:asymmetry_produced}, we present a numerical analysis covering a variety of initial conditions and scales for new physics, and we identify the allowed parameter space for a successful leptogenesis. 

\section{Initial Conditions for the Higgs VEV}
\label{sec:Higgs_IC}

We begin by motivating our project with the observation that the Higgs field can acquire a large vacuum expectation value (VEV) for a variety of reasons during inflation; therefore, an epoch of post-inflationary Higgs relaxation is a general feature of many cosmological scenarios.  In this work we are interested in generating an excess of leptons over antileptons during this epoch.  

When we generate the lepton asymmetry during this epoch of Higgs relaxation, we will find that the resulting asymmetry depends on the initial value of the VEV, denoted $\sqrt{\left<\phi^{2}\right>}=\phi_{0}$.  During inflation, quantum fluctuations of the Higgs field were ongoing, and therefore different patches of the Universe had slightly different VEVs at the end of inflation.  Regions that begin with slightly different $\phi_0$ values consequently develop different baryon asymmetries.  This produces unacceptably large baryonic isocurvature perturbations~\cite{Peebles1987,*1987ApJ...315L..73P,*Enqvist:1998pf,*Enqvist:1999hv,*Harigaya:2014tla}, which are constrained by CMB observations~\cite{Ade:2013uln}.  Therefore, in order to suppress isocurvature perturbations in the late universe, it is necessary to have a small variation in these values.  

In this section, we discuss two ways of generating the requisite large VEVs while suppressing the variation between different spacetime regions: through quantum fluctuations, which are suppressed by a Higgs-inflaton coupling until the end of inflation, and by trapping the Higgs field in a false vacuum.

The Standard Model Higgs boson has a tree-level potential 
\begin{equation}
V(\Phi) = m^2 \Phi^\dagger \Phi + \lambda (\Phi^\dagger \Phi)^2,
\end{equation}
where $\Phi$ is an $\mathrm{SU}(2)$ doublet.  The classical field may be written as
\begin{equation}
\Phi = \dfrac{1}{\sqrt{2}} \begin{pmatrix}
e^{i \theta} \phi \\ 0
\end{pmatrix},
\end{equation}
where $\phi(x)$ is a real scalar field.  The parameters $m$ and $\lambda$, although constant at tree-level, are modified by both loop and finite temperature corrections.  For the experimentally preferred top quark mass and Higgs boson mass, loop corrections result in a negative running coupling $\lambda$ at sufficiently large VEVs, with the result that the $\phi = v_\mathrm{EW} = 246 \; \mathrm{GeV}$ minimum is metastable at zero temperature~\cite{Degrassi:2012ry}.  We note, however, that a stable vacuum is possible within current experimental uncertainties~\cite{Degrassi:2012ry}. 
 
The running of the quartic coupling produces a shallow potential, which has the consequence that that a large VEV develops during inflation due to quantum fluctuations, at least in the regime in which the Standard Model vacuum is stable~\cite{Enqvist:2013kaa}.  We consider this sort of scenario in subsection IC-2 below.  Alternatively, the metastability of the electroweak vacuum is frequently possible within the inflationary paradigm~\cite{Kusenko:1996xt,*Kusenko:1996jn,*Kobakhidze:2013tn}, and the Higgs potential may be sensitive to higher-dimensional operators which lift the second minimum.  We consider this scenario in the subsequent subsection. 

\subsection{IC-1: Metastable Vacuum at Large VEVs}
\label{subsec:IC1}

At the large VEVs, the Higgs potential may be sensitive to the effects of higher-dimensional operators, which can lift the second minimum and consequently stabilize the electroweak vacuum.  The Higgs VEV may take an initial large value during inflation, similar to the initial VEV of the inflaton field itself in chaotic inflation models.  During inflation, such a VEV evolves towards the false vacuum from above, and then remains trapped in this false vacuum until destabilized by thermal corrections in reheating.  Subsequently, the field rolls to the global minimum at $\phi = 0$, until electroweak symmetry is broken at a significantly later time.

In order to lift the second minimum, we consider terms of the form
\begin{equation}
\mathcal{L}_\mathrm{lift} = \dfrac{\phi^{10}}{\Lambda_{\text{lift}}^{6}}.
\label{eq:lift}
\end{equation}
This non-renormalizable operator may be viewed as an effective operator arising from integrating out heavy states in loops.

During inflation, thermal corrections in the supercooled universe are insufficient to destabilize the metastable vacuum.  We also ensure that the quantum fluctuations (discussed in detail in the next subsection) do not destabilize the vacuum by requiring that the potential barrier height $\Delta V \gg H_I^4$.  In order to suppress the above-mentioned isocurvature perturbations, we will ensure that fluctuations about the false minimum are able to relax back to the minimum, for which it is sufficient to ensure $m_\mathrm{eff} \sim \sqrt{ d^2 V \slash d\phi^2} > H_I$ in the region probed by quantum fluctuations.

As a specific example, we consider the Higgs potential with one loop corrections~\cite{Degrassi:2012ry} with the experimentally preferred values $m_h = 126$~GeV and $m_t = 173.07$~GeV.  Taking $\Lambda_{\text{lift}} = 6.52 \times 10^{15}$~GeV gives a metastable minimum near $\phi=10^{15}$~GeV, with a potential barrier height of $\Delta V \approx 10^{53} \; \mathrm{GeV}^4$.  We will consider $H_I \sim 10^{11} \; \mathrm{GeV}$; in addition to being insufficient to probe the region beyond the barrier, this is less than the effective mass $m_\mathrm{eff} \sim 10^{13} \; \mathrm{GeV}$ in the region probed by quantum fluctuations.  Provided that the maximum reheat temperature is greater than $\sim 5 \times 10^{13} \; \mathrm{GeV}$, thermal corrections during reheating are sufficient to destabilize this vacuum.

\subsection{IC-2: Quantum Fluctuations}
\label{subsec:IC2}


The running coupling constant $\lambda$ results in a shallow potential, and during inflation, scalar fields with slowly rising potentials generically develop large VEVs.  Qualitatively, the scalar field in a de Sitter space can develop a large VEV via quantum effects, such as Hawking-Moss instantons~\cite{Bunch:1978yq,Hawking:1981fz} or stochastic growth~\cite{Linde:1982uu,Starobinsky:1982ee,Vilenkin:1982wt}.  The field then relaxes to its equilibrium value via a classical motion, which requires a time
\begin{equation}
\tau_{\phi}\sim m_{\mathrm{eff}}^{-1}\sim\left( \sqrt{d^{2}V\slash d\phi^{2}}\right) ^{-1}.
\end{equation}

If the universe expands sufficiently quickly during inflation, then relaxation is too slow and quantum jumps occur frequently enough to maintain a large VEV.  Specifically, large VEVs occur if the Hubble parameter $H_I = \sqrt{8\pi \slash 3} \Lambda_I^2 \slash M_\mathrm{Pl} \gg \tau_\phi^{-1}$.
For field values $\phi$ that satisfy this relation, Hubble friction is sufficient to prevent the system from relaxing to its equilibrium value $\phi = 0$.  Averaged over superhorizon scales, the mean Higgs VEV is such that $V(\phi_I) \sim H_I^4$~\cite{Bunch:1978yq,Hawking:1981fz,Enqvist:2013kaa}, provided that this VEV does not probe the second vacuum in the case that the electroweak vacuum is quasistable.  


Although the average vacuum expectation value is $\phi_I$, there is variation between the VEVs of different horizon-sized patches.  Consequently, different patches of the observable universe began with different $\phi_0$ values, and as discussed above, this generically results in unacceptably large isocurvature perturbations.  However, also as mentioned, the Higgs potential is sensitive to the effects of higher-dimensional operators at large VEVs; here we use such operators to limit the growth of the Higgs VEV to the last several e-folds of inflation.  This has the result that the isocurvature perturbations are limited to smaller angular resolution scales than have been experimentally probed.  Specifically, we introduce one or more couplings between the Higgs and inflaton field of the form
\begin{equation}
\mathcal{L}_\mathrm{\phi I} = c \dfrac{(\Phi^\dagger \Phi)^{m/2} (I^\dagger I)^{n \slash 2}}{M_\mathrm{Pl}^{m+n-2}},
\label{eq:inflaton_coupling}
\end{equation}
which increases the effective mass of the Higgs field during the early stages of inflation, when $\langle I \rangle $ is large (superplanckian, in the case of chaotic inflation).  As explained above, when $\tau_\phi^{-1} \sim m_\mathrm{eff}(\phi_I) \sim H_I$ the expansion of the universe is not sufficiently rapid to trap the field at large VEVs.  At the end of slow-roll inflation, $\left< I \right>$ decreases; consequently this term becomes negligible and the Higgs acquires a large vacuum expectation value.  

If the Higgs VEV grows during the last $N_\mathrm{last}$ e-folds of inflation, it reaches the average value 
\begin{equation}
\phi_0=\min [\phi_I, \sqrt{N_\mathrm{last}} H_I \slash 2 \pi].
\label{eq:v0_IC2}
\end{equation}
Provided $N_\mathrm{last}\approx 5-8$, the baryonic isocurvature perturbations develop only on the smallest angular scales which are not yet constrained.      

We emphasize that operators of the form \eqref{eq:inflaton_coupling} may be viewed as effective operators arising from integrating out heavy states in loops.  We note that the change in $\langle I \rangle$ during the slow-roll phase of inflation is model-dependent, and consequently the allowed range of parameters $c$, $m$, and $n$ differs from model to model.  This range may be quite narrow, and so this scenario may require some fine-tuning.

As a concrete example, we consider only the term,
\begin{equation}
V_\mathrm{mix} = \dfrac{1}{2} \dfrac{I^{2n}}{M^{2 n-2}} \phi^2,
\end{equation}
which induces an effective mass $m_\mathrm{eff}(\left< I \right>) = \left< I \right>^n \slash M^{n-1}$ for the Higgs field.  We define $I_1$ as the VEV of the inflaton field value at the end of slow roll inflation, and $I_2$ as the VEV of the inflaton field 8 e-folds before the end of slow roll inflation.  To ensure that the Higgs VEV grows only during the last e-folds, we must choose parameters such that $m_\mathrm{eff}(I_2) \approx H_I$.  We illustrate this approach with quartic inflation (although this is disfavored observationally; see Ref.~\cite{Ade:2013uln}).  With the inflaton potential $V_I = \lambda_I I^4$, slow roll inflation ends when the inflaton as a vacuum expectation value of $I_1 = M_\mathrm{Pl} \slash \sqrt{2\pi}$.  The number of e-folds during the time in which the inflaton evolves from $\left<I \right>$ to $I_1$ is
\begin{align}
N(\left< I \right> \rightarrow I_1) = \pi \left( \dfrac{\left< I \right>}{M_\mathrm{Pl}} \right)^2 - \dfrac{1}{2},
\end{align}
which gives
\begin{equation}
I_2 = \sqrt{ \dfrac{17}{2\pi} } M_\mathrm{Pl}.
\end{equation}
(Although this field value is superplanckian, this is a feature of quartic inflation which does not necessarily apply to other inflationary models.)  The Hubble parameter at this field value is given by
\begin{equation}
H_I^2(I_2) = \dfrac{8\pi}{3 M_\mathrm{Pl}^2} \lambda_I I_2^4 = \dfrac{8\pi}{3} \left( \dfrac{17}{2\pi} \right)^2 \lambda_I M_\mathrm{Pl}^2.
\end{equation}
The quartic coupling $\lambda_I$ must be $\lesssim 10^{-13}$ in order to avoid large CMB temperature anisotropies, which gives $M \sim 10^6 M_\mathrm{Pl}$ for both $n=2$ and $n=4$.  In this way, a coupling between the Higgs field and the inflaton field can prevent the Higgs VEV from growing until the last several $N$-folds of inflation, suppressing the scale of isocurvature perturbations.  As this example illustrates, the constraints on the Higgs-inflaton coupling depends on the shape of the inflaton potential.  Although we have demonstrated an explicit calculation using a quartic inflationary potential, a similar calculation can be done with other potentials.

Our analysis of the final asymmetry will depend only on the VEV of the Higgs field at the end of inflation, which as noted above is $\phi_0=\min [\phi_I, \sqrt{N_\mathrm{last}} H_I \slash 2 \pi]$ provided that the parameters in whatever inflationary model is used are chosen such that the Higgs VEV does not begin to grow until the last $N_\mathrm{last}$-folds of inflation.  Therefore, we do not specify a specific inflationary model in our analysis, and we take the inflationary scale $\Lambda_n$ and the inflaton decay rate $\Gamma_I$ to be free parameters.

\section{Reheating}
\label{sec:Reheat}

Now that we have established that the Higgs field can develop a large VEV during inflation, we are interested in its subsequent evolution to its equilibrium value.  Relaxation begins when the Hubble parameter is comparable to the effective mass of the Higgs field, $m_\mathrm{eff}(\phi) \approx H(I)$, which is within the reheating epoch.  Therefore, this relaxation is sensitive to finite temperature effects due to the plasma, and so we now proceed to discuss reheating.


In both scenarios, whether IC-1 or IC-2, we ensure that the energy density is never dominated by the Higgs field.  Inflaton oscillations dominate until the transition to the radiation dominated era, which occurs when the inflaton decay width is comparable to the Hubble parameter, $\Gamma_I \sim H_\mathrm{RH}$, which typically occurs after the Higgs field has lost a significant portion of its energy.  Consequently, the reheat temperature $T_\mathrm{RH} \sim \sqrt{ \Gamma_I M_\mathrm{Pl}}$, is generally only weakly constrained~\cite{Dai:2014jja}.

For simplicity, we assume coherent oscillations begin instantly at the end of the inflationary epoch, and as a simple model, we assume that the inflaton decays entirely to radiation at a constant rate,
\begin{equation}
\dot{\rho}_r  + 4 H(t) \rho_r = \Gamma_I \rho_I,
\end{equation}
where
\begin{equation}
\rho_I = \dfrac{\Lambda_I^4 e^{-\Gamma_I t}}{a(t)^3}
\end{equation}
is the energy density of the inflaton field.  The evolution of the Hubble parameter is given by
\begin{equation}
H(t) \equiv \dfrac{\dot{a}}{a} = \sqrt{ \dfrac{8 \pi}{3 M_\mathrm{Pl}^2} (\rho_r + \rho_I)  }.
\end{equation}
This is a complete system of equations that may be solved independently of the evolution of the Higgs condensate.  Throughout this work, we take $t=0$ to be the beginning of the coherent oscillation of the inflaton field; during the coherent oscillation epoch, the universe evolves as if it were matter dominated, until the radiation from reheating dominates.

During reheating, the effective temperature of the plasma is defined using the radiation density as
\begin{equation}
\rho_{r}=\frac{g_{*}\pi^{2}}{30}T^{4}.
\label{eq:rho_R}
\end{equation}
For $t \gg t_i = (2 \slash 3) \sqrt{ 3 \slash 8 \pi} M_{{\rm Pl}} \slash \Lambda_I^2$, the temperature evolves as
\begin{equation}
T=\left( \frac{3}{g_*\pi^3}\frac{\Gamma_I M_\mathrm{Pl}^2}{t}\right)^{1/4},
\label{eq:time1}
\end{equation}
until it reaches the reheat temperature $T_R\sim \sqrt{\Gamma_I M_\mathrm{Pl}}$.  Subsequently, radiation dominates the energy density and the temperature evolves as 
\begin{equation}
 T=\left( \frac{45}{16\pi^3 g_*}\right)^{1/4} \sqrt{M_\mathrm{Pl}/t} .
 \label{eq:time2}
\end{equation}

\section{Evolution of the Higgs VEV}
\label{sec:Relaxation}

We now turn our attention to the relaxation of the Higgs VEV, which evolves as~\cite{Kolb:1990vq}
\begin{equation}
\ddot{\phi} + 3 H(t) \dot{\phi} + V_\phi^\prime (\phi,T(t)) + \Gamma_{H}\dot{\phi} = 0, 
\label{eq:Higgs_VEV_eqn_motion}
\end{equation}
where $V_\phi(\phi,T)$ is the Higgs effective potential, including modifications from the decays of the condensate~\cite{Enqvist:2013kaa}.  $\Gamma_{H}$ describes the effect of the perturbative decay of the condensate.  In the first subsection below, we discuss the one-loop corrected potential, including one-loop corrections to the RG equations.  Subsequently, we consider the non-perturbative decay of the Higgs condensate, followed by perturbative decay.  Finally, we present a numerical analysis of the evolution of the Higgs condensate, before we proceed to the next section which introduces the relevant higher dimensional operator we use to produce the nonzero lepton asymmetry.

\subsection{Effective Potential}

The Standard Model Higgs potential computed to a fixed order in perturbation theory is generally gauge-dependent, although the value of the potential at the extrema are not (see, for example, \cite{Andreassen:2014gha,Andreassen:2014eha}).  One can ensure gauge-invariant results by removing the gauge-dependence of the potential using Nielsen identities~\cite{Nielsen:1975fs,Fukuda:1975di,Aitchison:1983ns}.  Here we use the Landau gauge, which has good numerical agreement with the corrected potential~\cite{Andreassen:2014gha,DiLuzio:2014bua}.  In our analysis, we have used the one-loop corrected potential~\cite{Casas:1994qy}, with running couplings (including one-loop corrections to the renormalization group (RG) equations, as given in~\cite{Degrassi:2012ry}).  The one-loop potential is 
\begin{widetext}
\begin{align}
V_\phi^\mathrm{1-loop} &= \dfrac{1}{2} m_\phi^2 \phi^2 + \dfrac{\lambda}{4} \phi^4 + \dfrac{1}{(4\pi)^2} \left[ \dfrac{m_H(\phi)^4}{4} \left( \ln \left( \dfrac{m_H(\phi)^2}{\mu^2} \right) - \dfrac{3}{2} \right) 
+ \dfrac{3 m_G(\phi)^4}{4} \left( \ln \left( \dfrac{m_G(\phi)^2}{\mu^2} \right) - \dfrac{3}{2} \right) \right. \nonumber \\
&\left. + \dfrac{3 m_W(\phi)^4}{2} \left( \ln \left( \dfrac{m_W(\phi)^2}{\mu^2} \right) - \dfrac{5}{6} \right)
+ \dfrac{3 m_Z(\phi)^4}{4} \left( \ln \left( \dfrac{m_Z(\phi)^2}{\mu^2} \right) - \dfrac{5}{6} \right)
- 3 m_t(\phi)^4 \left( \ln \left( \dfrac{m_t(\phi)^2}{\mu^2} \right) - \dfrac{3}{2} \right)
\right],
\end{align}
\end{widetext}
where $\mu$ is the renormalization scale and the tree-level masses for the Higgs boson, Goldstone mode, $W$ bosons, $Z$ boson, and top quark are
\begin{alignat}{3}
m_W^2 &= \dfrac{g^{2} \phi^2}{4}, &\quad m_Z^2 &= \dfrac{(g^2 + g^{\prime \, 2}) \phi^2}{4}, \quad &
m_t &= \dfrac{y_t \phi}{\sqrt{2}}, \nonumber \\
m_H^2 &= m_\phi^2 + 3 \lambda  \phi^2, &\quad m_G^2 &= m_\phi^2 + \lambda.\label{eq:Tree-level mass}
\end{alignat}

We have also included the finite temperature corrections~\cite{Anderson:1991zb,Kapusta:2006pm},
\begin{align}
V_T(\phi,T) &= -\dfrac{T^2}{2 \pi^2} \left[ 6 m_W^2 J_B \left( \dfrac{m_W}{T} \right) + 3 m_Z^2 J_B \left( \dfrac{m_Z}{T} \right) \right.  \nonumber \\
& \left. \qquad + 12 m_t^2 J_F \left( \dfrac{m_t}{T} \right) \right],
\end{align}
where
\begin{align}
J_{B}(y) & =\sum_{n=1}^{\infty}\frac{1}{n^{2}}K_{2}(ny),\\
J_{F}(y) & =\sum_{n=1}^{\infty}\frac{(-1)^{n+1}}{n^{2}}K_{2}(ny),
\end{align}
and we have ignored the contributions from Higgs bosons and Goldstone mode, which only dominate when $\phi\lesssim v_{\mathrm{EW}}$. We emphasize that we do not use the high temperature expansion, as during reheating the condition $T(t) \gg \phi(t)$ is not satisfied at all times.  The renormalization scale $\mu$ is taken to be $\sqrt{\phi^2 + T^2}$. 

We note that two-loop corrections may be significant at the boundary of the metastability region~\cite{Degrassi:2012ry}; however, a self-consistent analysis at two-loop order would include finite temperature effects in the RG equations, which is beyond the scope of this work.

After the Higgs VEV passes through zero, it generally oscillates around its minimum at $\phi=0$, which remains a minimum for $T\gg v_{\mathrm{EW}}$. During this oscillation, the Higgs condensate can then decay perturbatively and non-perturbatively into Standard Model particles. The non-perturbative decay happens much faster than the perturbative decay and is the dominant channel, as pointed out by \cite{Enqvist:2013kaa}.  We now proceed to discuss the effect of these decays.

\subsection{Non-Perturbative Decay}

First, we consider non-perturbative decay.  The oscillation of the Higgs field provides a time-dependent mass term for all the coupled particles, which can cause resonant production of the particles. The produced particles then induce an effective mass term to the Higgs condensate as a backreaction; this attenuates the oscillation of the Higgs field until the resonant production is off \cite{Enqvist:2013kaa,Enqvist:2014tta}.

The non-perturbative decay channel of Higgs is dominated by $h\rightarrow WW,\: ZZ$. The Lagrangian containing the Standard Model weak gauge fields and the Higgs sector is 
\begin{align}
\mathcal{L} & =\frac{1}{2}g^{\mu\nu}\partial_{\mu}\phi\partial_{\nu}\phi-V_{\phi}(\phi,T)+g^{\mu\nu}\left[\frac{1}{4}g^{2}W_{\mu}^{+}W_{\nu}^{-}\right.\nonumber \\
 & \;\left.+\frac{1}{8}\left(g^{2}+g'^{2}\right)Z_{\mu}Z_{\nu}\right]\phi^{2}+\mathcal{L}_{A\text{, kin}},
\end{align}
where the kinetic terms of the gauge fields can be expanded as
\begin{align}
\mathcal{L}_{A\text{, kin}} & =-\frac{1}{2}\left(\nabla_{\mu}W_{\nu}^{+}-\nabla_{\nu}W_{\mu}^{+}\right)\left(\nabla^{\mu}W^{\nu-}-\nabla^{\nu}W^{\nu-}\right)\nonumber \\
 & -\frac{1}{4}\left(\nabla_{\mu}Z_{\nu}-\nabla_{\nu}Z_{\mu}\right)^{2}+O(g)(\text{non-Abelian terms}).
\end{align}
Since the non-Abelian contributions are small at the beginning of the resonant production of $W$ and $Z$ bosons, we ignore those terms~\cite{Enqvist:2013kaa,Enqvist:2014tta}. We also work specifically in flat FLRW spacetime, $g_{\mu\nu}=a^{2}\left(\tau\right)\eta_{\mu\nu}$, with conformal time $\tau=\int a^{-1}dt$. The resonant production of the weak gauge fields, $A_{\mu}=W_{\mu}^{\pm}\:\text{or}\: Z_{\mu}$, in momentum space is then described by 
\begin{equation}
A_{0}\left(\vec{k},\tau\right)=\frac{-ikA'_{L}\left(\vec{k},\tau\right)}{k^{2}+a^{2}m_{A}^{2}\left(\phi\right)},
\label{eq:A0}
\end{equation}
\begin{equation}
A''_{T,i}+\omega_{k}^{2}\left(\phi\right)A_{T,i}=0,
\label{eq:AT}
\end{equation}
and 
\begin{equation}
A''_{L}+\omega_{k}^{2}\left(\phi\right)A_{L}+\frac{2k^{2}}{\omega_{k}^{2}\left(\phi\right)}\partial_{\tau}\ln\left(am_{A}\right)A'_{L}=0,
\label{eq:AL}
\end{equation}
where $\omega_{k}=\sqrt{k^{2}+a^{2}m_{A}^{2}\left(\phi\right)}$ and prime denotes differentiation with respect to conformal time $d\tau$. $\vec{A}_{T}\left(\vec{k},t\right)$ and $A_{L}\left(\vec{k},t\right)$ are the transverse and longitudinal components of the spatial component $\vec{A}\left(\vec{k},t\right)$, respectively. The mass term $m_{A}^{2}\left(\phi\right)$ is given in Eq.\ \eqref{eq:Tree-level mass} and can include the thermal correction by replacing $\phi^{2}\rightarrow\phi^{2}+C_{A}T^{2}$ where we use $C_{W}=2/3$, and $C_Z < 1$ is determined by diagonalizing the mass matrix~\cite{Elmfors:1993re}. Due to extra friction term in Eq.~\eqref{eq:AL} for the longitudinal component $A_{L}$, we expect the resonance production of this mode to be suppressed. $A_{0}$, which depends only on $A_{L}$ through Eq.\ \eqref{eq:A0}, should also be suppressed \cite{Enqvist:2014tta}. Hence, we focus on the transverse mode $A_{T}$ only. 

Resonant production of particles can be understood as the amplification of vacuum fluctuations. The number of particles in each mode produced from the vacuum is
\begin{equation}
n_{k}=\frac{1}{2\omega_{k}}\left(\left|A'_{\mu}(\vec{k},\tau)\right|^{2}+\omega_{k}^{2}\left|A_{\mu}(\vec{k},\tau)\right|^{2}\right)-\frac{1}{2}.
\end{equation}
The initial conditions are taken to be the WKB approximation of the vacuum solution,
\begin{equation}
A_{T}(k,0)=\frac{1}{\sqrt{2\omega_{k}}};\quad A'_{T}(k,0)=-i\sqrt{\frac{\omega_{k}}{2}},
\end{equation}
which satisfy $n_{k}(0)=0$ and the Wronskian condition $AA'^{*}-A^{*}A'=i$.

\begin{figure}
\includegraphics[width=1\columnwidth]{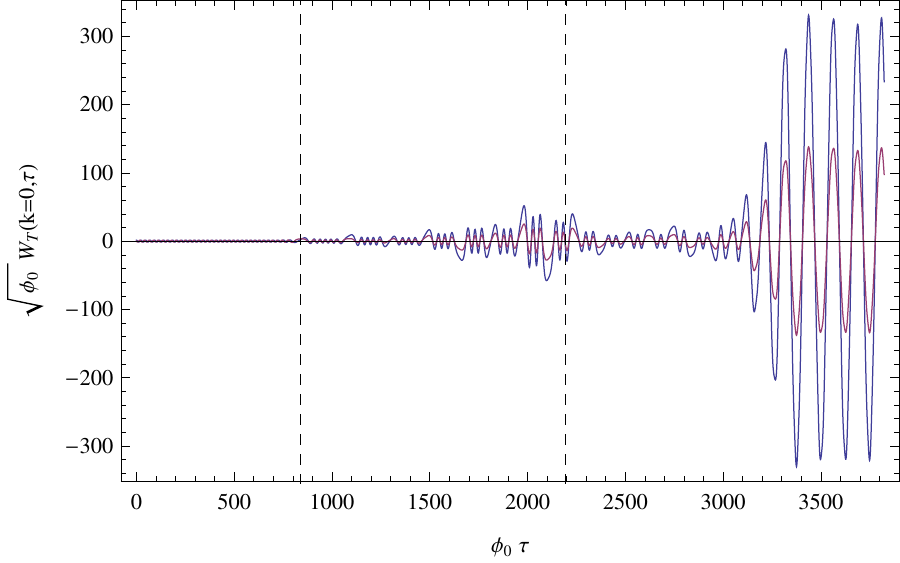}
\caption{Real and imaginary parts (blue and purple lines) of $W_{T}(\tau)$ for $k=0$ for IC-1, with the parameters $\Lambda_{I}=10^{15}\;\mathrm{GeV}$ and $\Gamma_{I}=10^{9}\;\mathrm{GeV}$. The vertical lines designate the first time the Higgs VEV crosses zero, and the time of maximum reheating, from left to right.}
\label{fig:W_prod}
\end{figure}

\begin{figure}
\includegraphics[width=1\columnwidth]{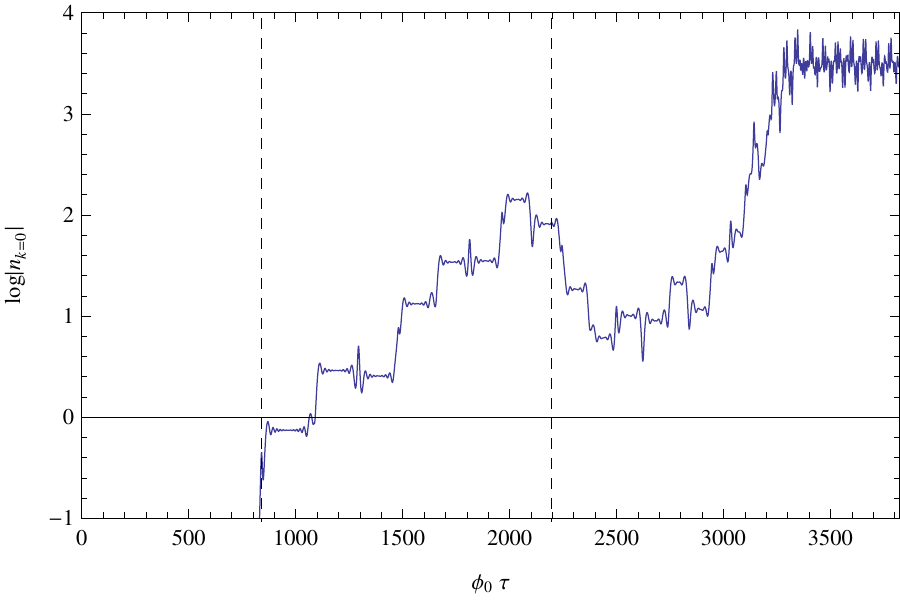}\protect\caption{ $\text{log}\left[n_{k}(\tau)\right]$ for $k=0$, with the same parameters as Fig.\ \ref{fig:W_prod}. Note $n_{k}\left(\tau\right)$ stops increasing at $\tau\sim3300/\phi_{0}$ because the amplitude of $\phi$ has decreased to the order of $T$. The effective mass of $W$ was then dominated by $T$ instead of $\phi$.
\label{fig:nk0}}
\end{figure}

Fig.\ \ref{fig:W_prod} shows the amplification of the $W$ field, which increases each time the Higgs VEV passes through zero. The number density $n_{k}$ is shown in Fig.~\ref{fig:nk0}. It has a sequence of flat steps, which are separated by peaks. Those peaks occur when $\phi=0$; due to the rapidly changing mass, the number of particles is not well defined at these points. Particle number is well defined only when $\phi$ reaches a local maxima or minima. We approximate the particle number of $A_{\mu}$ quanta within each oscillation of $\phi$ by its value when $\dot{\phi}=0$, which is supported by the flatness of the steps in Fig.~\ref{fig:nk0}. The resonant production begins once the Higgs VEV starts to oscillate at $\tau\sim800/\phi_{0}$. The decrease of $n_{k}$ at $\tau\sim2500/\phi_{0}$ indicates the system has a stochastic resonance, which is a distinctive feature of parametric resonance in an expanding universe \cite{Kofman:1997yn}. The resonant production then ceases at $\tau\sim3300/\phi_{0}$ because the amplitude of $\phi$ has decreased to the order of $T$.


If we approximate the oscillation of the Higgs VEV by $\phi(\tau)=\phi_{m}\cos(\omega_{\phi}\tau)$,
we can write Eq.\ \eqref{eq:AT} as a Mathieu equation of the form
\begin{equation}
\frac{d^{2}A_{T}}{dz^{2}}+\left(m^{2}+b^{2}\cos^{2}z\right)A_{T}=0
\end{equation}
where $z=\omega_{\phi}\tau$, $m^{2}\approx\left[k^{2}+a^{2}m_{A}^{2}\left(\phi=0,T\right)\right]/\omega_{\phi}^{2}$ and $b^{2}\approx a^{2}m_{A}^{2}\left(\phi_{m},T=0\right)/\omega_{\phi}^{2}$. The Mathieu equation has an instability only when $b\gtrsim m^{2}$. Thus, resonant production is suppressed for 
\begin{equation}
k>k_\mathrm{max}\approx\sqrt{a\omega_{\phi}m_{A}\left(\phi_{m},0\right)-a^{2}m_{A}^{2}\left(0,T\right)}.
\end{equation}

The produced $W$ and $Z$ fields induce an effective mass for the Higgs field as a backreaction,
\begin{align}
m_{\phi,W}^{2} & =-\frac{1}{2}g^{2}\left\langle W_{\mu}^{+}W^{\mu-}\right\rangle \label{eq:mHw}\\
m_{\phi,Z}^{2} & =-\frac{1}{4}\left(g^{2}+g'^{2}\right)\left\langle Z_{\mu}Z^{\mu}\right\rangle \label{eq:mHz}
\end{align}
where the expectation value of $A=W,\: Z$ can be approximated as
\cite{Kofman:1997yn,Enqvist:2013kaa}
\begin{equation}
g^{\mu\nu}\left\langle A_{\mu}A_{\nu}\right\rangle \cong\frac{-2}{a^{2}}\left\langle A_{T}^{2}\right\rangle \approx\frac{-1}{\pi^{2}a^{2}}\int_{0}^{\infty}\frac{k^{2}dk}{\omega_{k}}n_{k}.\label{eq:AAexpect}
\end{equation}
In general, the integral in Eq.\ \eqref{eq:AAexpect} will need to be regularized.  However, in our case there is no significant contribution from $n_{k}$ values with $k\gtrsim k_{max}$, and so the integral is finite. The upper limit can be approximated by $k_{max}$. One can then include the non-perturbative decay of the Higgs by adding the induced mass terms Eqs.\ \eqref{eq:mHw} and \eqref{eq:mHz} into the Higgs potential in Eq.\ \eqref{eq:Higgs_VEV_eqn_motion}.

\begin{figure}
\includegraphics[width=1\columnwidth]{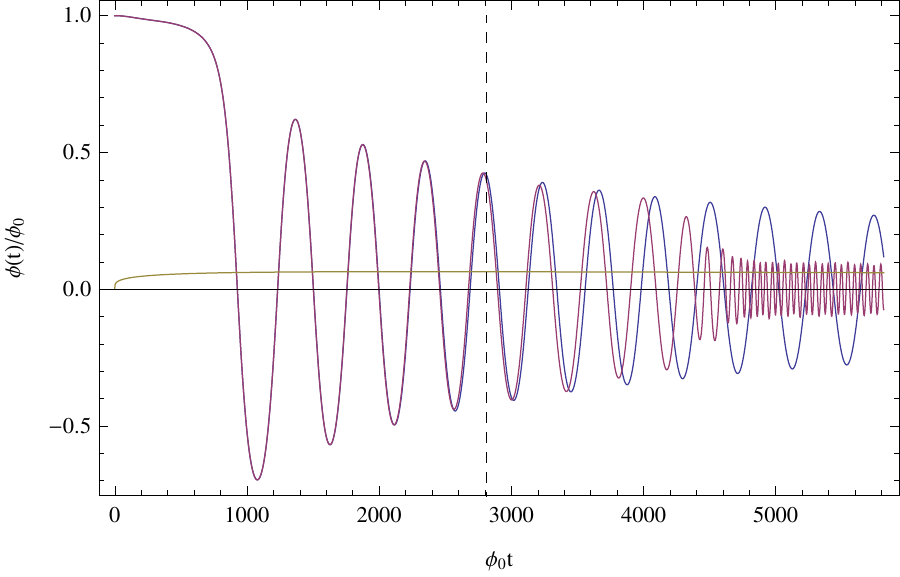}\protect\caption{Non-perturbative decay of the Higgs condensate for IC-1, with the parameters $\Lambda_{I}=10^{15}\;\mathrm{GeV}$ and $\Gamma_{I}=10^{9}\;\mathrm{GeV}$. The purple (blue) line corresponds to evolution of the Higgs VEV with (without) non-perturbative decay. The brown line corresponds to the temperature of the plasma. The vertical dashed line indicates the time of maximum reheating.}
\label{Higgs VEV with non-perturbative decay}
\end{figure}

Fig.\ \ref{Higgs VEV with non-perturbative decay} shows an example of the Higgs evolution with the non-perturbative decay for the IC-1 scenario. The increasing effective masses from $W$ and $Z$ affect the oscillation of Higgs when $m_{\phi,A}^{2}\gtrsim T$; these decrease the amplitude of the Higgs oscillation. When the Higgs VEV decreases to $\phi\lesssim T$, the resonant production of $W$ and $Z$ end, because the non-perturbative decay channel is blocked by the large $W$ and $Z$ thermal masses. In this case, one has only to consider perturbative decay channels, discussed in subsection \ref{sub:Perturbative-Decay-Thermalization}.

Note the generated $W$ and $Z$ bosons can decay perturbatively into fermions. This decay could in principle obstruct the resonant production of $W$ and $Z$ in the usual Standard Model case~\cite{GarciaBellido:2008ab,Figueroa:2015rqa}. However, in the parameter space that we are interested in, the average decay times of $W$ and $Z$ bosons $\left\langle \Gamma_{W,\, Z}\right\rangle ^{-1}$ are longer than the semiperiod of the Higgs oscillation. Thus, we have ignored the decay of $W$ and $Z$ in our analysis. 

The analysis in this subsection can be improved by using a lattice gauge theory~\cite{GarciaBellido:2003wd,Figueroa:2015rqa}.  However, as Fig.~\ref{Higgs VEV with non-perturbative decay} demonstrates, the non-perturbative decay of the Higgs condensate is relevant only after several oscillations, whereas in our scenario the asymmetry will be generated primarily during the initial oscillation of the Higgs VEV.

\subsection{Perturbative Decay - Thermalization\label{sub:Perturbative-Decay-Thermalization}}

The perturbative decay of the Higgs condensate is described by the friction term $\Gamma_{H}\dot{\phi}$ in the equation of motion \eqref{eq:Higgs_VEV_eqn_motion}.  The decay width can be computed through the imaginary part of the self-energy operator
\begin{equation}
\Gamma_{H}=\frac{\text{Im}\Pi}{m_{\mathrm{eff}}},
\end{equation}
where $m_{\mathrm{eff}}=\text{Re}\sqrt{\partial^{2}V_{\phi}\left(\phi,T\right) \slash \partial \phi^2}$ is the effective mass of the Higgs boson. In a finite-temperature thermal background,  $\Gamma_{H}$ corresponds to the thermalization rate of the Higgs condensate. 

Here we consider the fermionic decay channels, motivated by the large top Yukawa coupling.  (The dominant bosonic channels, $WW$ and $ZZ$, are included in the non-perturbative calculation, which dominates their perturbative contribution.)  In the thermal bath of fermions, there are additional excitations which are the removals of antiparticles from the Fermi sea (holes). The dispersion relations for particles and holes are \cite{Weldon:1989ys,Elmfors:1993re,Enqvist:2004pr}
\begin{align}
\hat{\omega}_{p}-\hat{k}-\frac{g_{T}^{2}}{\hat{k}}-\frac{g_{T}^{2}}{2\hat{k}}\left(1-\frac{\hat{\omega}_{p}}{\hat{k}}\right)\ln\left|\frac{\hat{\omega}_{p}+\hat{k}}{\hat{\omega}_{p}-\hat{k}}\right| & =0,\label{eq:Ep}\\
\hat{\omega}_{h}+\hat{k}+\frac{g_{T}^{2}}{\hat{k}}-\frac{g_{T}^{2}}{2\hat{k}}\left(1+\frac{\hat{\omega}_{h}}{\hat{k}}\right)\ln\left|\frac{\hat{\omega}_{h}+\hat{k}}{\hat{\omega}_{h}-\hat{k}}\right| & =0,\label{eq:Eh}
\end{align}
where $\hat{\omega}\left(k\right)=\omega/T$, $\hat{k}=k/T$, and the subscripts $p$ and $h$ refer to particles and holes respectively. Eq.\ \eqref{eq:Eh} can also be expressed as
\begin{equation}
\hat{\omega}_{h}=\hat{k}\coth\left(\frac{\hat{k}^{2}}{g_{T}^{2}}+\frac{\hat{k}}{\hat{\omega}_{h}+\hat{k}}\right),\label{eq:Eh coth}
\end{equation}
which is a convenient form for numerical purposes.  We will specify the necessary coefficient $g_T$ below.  In these equations, we have made the approximation that the left- and right-handed fermions have the same thermal mass $m\left(T\right)=g_{T}T$; generically, this is not true because they are in different representations of the Standard Model gauge group.  However, this difference, which is much smaller than the difference between the particle and hole contributions, is negligible~\cite{Elmfors:1993re}.

The dominant fermionic contribution to the thermalization of the Higgs condensate is from the top quark, due to the large top-Higgs Yukawa coupling. The thermal mass of the left-handed top quark is \cite{Elmfors:1993re} 
\begin{equation}
g_{T,t}=\sqrt{\frac{1}{6}g_{s}^{2}+\frac{3M_{W}^{2}+\frac{1}{9}\left(M_{Z}^{2}-M_{W}^{2}\right)+M_{t}^{2}+M_{b}^{2}}{8v_{\mathrm{EW}}^{2}}},\label{eq:gT mass}
\end{equation}
where $M_{i}$ are the physical masses at $T=0$, and the strong coupling $g_{s}\cong1.220$. 

The presence of particles and holes in the fermionic plasma provides two thermalization processes for the Higgs condensate. A Higgs boson can decay into a pair of particles or a pair of holes respectively if
\begin{equation}
m_{\text{eff}}=2\omega_{i}\left(k_{i}\right);\quad i=p,h
\end{equation}
is satisfied.  The contribution of each process to the decay width is
\begin{equation}
\frac{\text{Im}\Pi_{\text{dec}}}{T^{2}}=\frac{y_t^{2}}{4\pi g_{T}^{4}}\sum_{i=p,h}\hat{k}_{i}^{2}\left(\hat{\omega}_{i}^{2}-\hat{k}_{i}^{2}\right)^{2}\left(1-2n_{i}\right)
,\end{equation}
where $y_t$ is the top Yukawa coupling, and 
\begin{equation}
n_{h,p}=\frac{1}{\exp\left(\hat{\omega}_{h,p}\right)+1}
\end{equation}
are the fermion distribution functions. Although this decay channel is blocked when $m_{\phi}<2\min\left[\omega_{h}\left(k\right)\right]$, a Higgs boson can also be absorbed by a hole to produce a particle. The contribution of this absorption channel to the width is 
\begin{equation}
\frac{\text{Im}\Pi_{\text{abs}}}{T^{2}}=\frac{y_t^{2}}{2\pi g_{T}^{4}}\sum_{i}\hat{k}_{i}^{2}\left(\hat{\omega}_{p}^{2}-\hat{k}_{i}^{2}\right)\left(\hat{\omega}_{h}^{2}-\hat{k}_{i}^{2}\right)\left(n_{h}-n_{p}\right)
\end{equation}
where the index $i$ sums over the solutions of 
\begin{equation}
m_{\text{eff}}+\omega_{h}\left(k_{i}\right)=\omega_{p}\left(k_{i}\right).
\end{equation}
The total thermalization rate is then the sum of two channels $\text{Im}\Pi=\text{Im}\Pi_{\text{abs}}+\text{Im}\Pi_{\text{dec}}$.

\begin{figure}
\includegraphics[width=1\columnwidth]{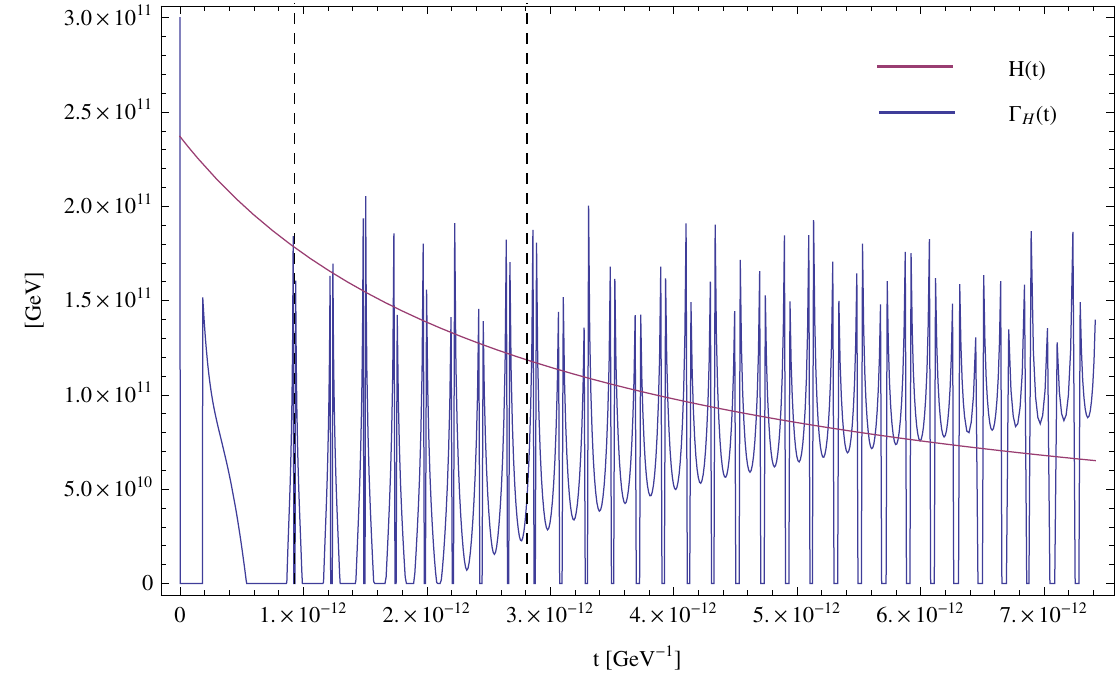}

\protect\caption{Higgs thermalization rate through the top quark compared with the Hubble parameter for IC-1, with the parameters $\Lambda_{I}=10^{15}\;\mathrm{GeV}$ and $\Gamma_{I}=10^{9}\;\mathrm{GeV}$. The vertical lines designate the first time the Higgs VEV crosses zero, and the time of maximum reheating, from left to right.}
\label{Higgs thermalization rate vs Hubble}
\end{figure}

For IC-1, Fig.\ \ref{Higgs thermalization rate vs Hubble} shows the thermalization rate of the Higgs condensate through the top quark compared with the Hubble parameter. We see the thermalization rate is comparable to the Hubble parameter only after the maximum reheating has been reached. Therefore, the evolution of the Higgs VEV is affected only at the end of reheating later time of reheating as shown in Fig.\ \ref{Higgs thermalization rate vs Hubble}. 

We have repeated the above analysis with the bottom quark in place of the top quark and verified numerically that its contribution is negligible; we also note that plasma effects can delay thermalization~\cite{Drewes:2013iaa}.  We also remark that particularly for IC-2, the thermalization rate is frequently much smaller than the Hubble parameter.  Since thermalization occurs on such long time scales, it has no effect on our analysis.

\subsection{Numerical Results}

\begin{figure}
\includegraphics[width=1\columnwidth]{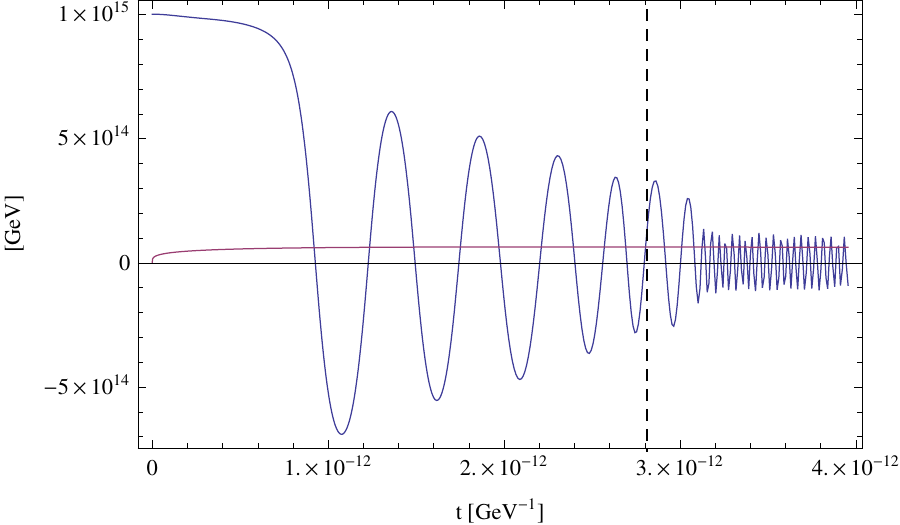}
\protect\caption{The evolution of the Higgs VEV for IC-1 (blue line) and temperature
(purple line) as a function of time, with the parameters $\Lambda_{I}=10^{15}\;\mathrm{GeV}$
and $\Gamma_{I}=10^{9}\;\mathrm{GeV}$. This plot includes both
the effect of non-perturbative decay and thermalization. The vertical line designate the time of maximum reheating.}
\label{fig:Higgs_Evolution_IC1} 
\end{figure}

\begin{figure}
\includegraphics[width=1\columnwidth]{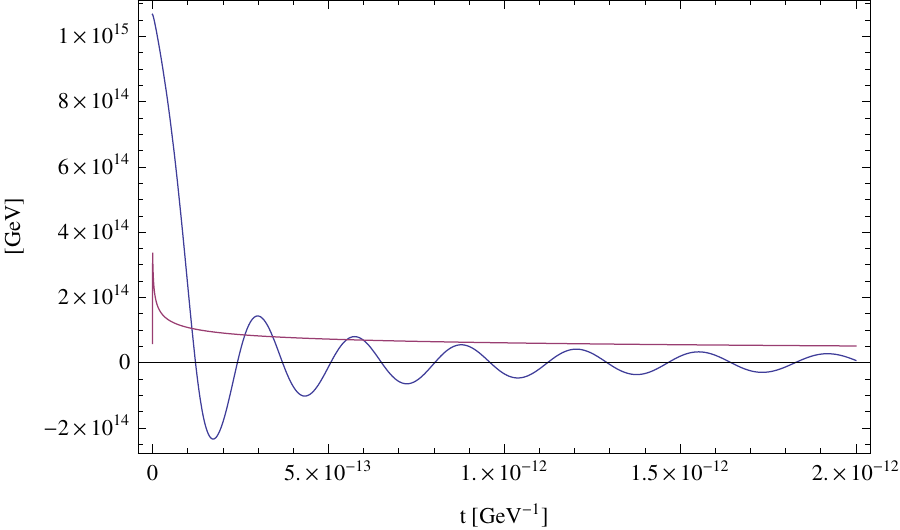}
\protect\caption{The evolution of the Higgs VEV for IC-2 (blue line) and temperature
(purple line) as a function of time, with the parameters $\Lambda_{I}=10^{17}\;\mathrm{GeV}$,
$\Gamma_{I}=10^{8}\;\mathrm{GeV}$, and $N_{\mathrm{last}}=8$.  This plot includes both
the effect of non-perturbative decay and thermalization, although
the effect of condensate decay is not appreciable in this case.}
\label{fig:Higgs_Evolution_IC2} 
\end{figure}

Figures \ref{fig:Higgs_Evolution_IC1} and \ref{fig:Higgs_Evolution_IC2} illustrate the evolution of the Higgs VEV (and temperature) as functions of time.  For IC-2, the relevant inflaton parameters are $\Lambda_I = 10^{17} \; \mathrm{GeV}$ and $\Gamma_I = 10^8 \; \mathrm{GeV}$, and we have assumed $N_\mathrm{last} = 8$ to determine $\phi_0$, which does not probe the quasistable vacuum.  For IC-1, we have used the operator (with the numerical parameters) discussed in section \ref{subsec:IC1} to lift the second minimum, along with the inflationary parameters $\Lambda_{I}=10^{15}\;\mathrm{GeV}$
and $\Gamma_{I}=10^{9}\;\mathrm{GeV}$.  For both plots, we use 126 GeV and 173.07 GeV for the masses of the Higgs boson and top quark respectively.  

We briefly remark on the qualitative features of these plots.  Although in IC-2, the Higgs VEV is constrained to grow only in the last 8 $N$-folds of inflation, it may still reach a large value if $H_I$ is large; for these inflationary parameters, $H_I = 2 \times 10^{15}$~GeV.  For other choices of $\Lambda_I$, the initial VEV $\phi_0$ for the IC-2 scenarios is significantly smaller.  Conversely, in our IC-1 scenario, the initial VEV $\phi_0 \approx 10^{15}$~GeV is set by the parameters chosen in section \ref{subsec:IC1}.

In scenario IC-1, the Higgs VEV remains approximately constant at short times, until reheating is sufficient to destabilize the second minimum, whereas in IC-2 the field relaxes as soon as the Hubble parameter becomes sufficiently small.  Subsequent oscillations have a larger amplitude in the IC-1 scenario; this is due to the difference in $\Lambda_I$ values, which result in less Hubble friction, and that the additional term \eqref{eq:lift} contributes to the velocity of the VEV.

A notable feature in \ref{fig:Higgs_Evolution_IC1} is that shortly after maximum reheating, the Higgs condensate begins to oscillate more rapidly.  This is due to the non-perturbative decay of the Higgs condensate, as illustrated in Fig.~\ref{Higgs VEV with non-perturbative decay} above.  (In the IC-2 scenario, such features are not relevant due to the rapid decay of the amplitude of oscillation.)

We see that both the thermal decay of the condensate and the non-perturbative decay of the condensate have little effect on first approach of the VEV to zero; in the scenario we outline below, the lepton asymmetry is generated primarily during this swing, and therefore, these processes have little effect on the total asymmetry generated.

\section{Effective Chemical Potential}
\label{sec:chemical_potential}

In the Introduction, we observed that the Higgs potential may be sensitive to the effects of higher dimensional operators, which are normally suppressed by powers of a high scale.  In section \ref{sec:Higgs_IC}, we have seen how such operators can be used to make a quasistable minimum in the Higgs potential or to suppress the growth of the Higgs VEV until the end stages of inflation.  Now, we consider an operator, involving only Standard Model fields, which generates an effective external chemical potential for leptons (and also baryons).  This operator is
\begin{equation}
\mathcal{O}_6 = - \dfrac{1}{\Lambda_n^2} \phi^2 \partial_\mu j^\mu,
\label{eq:O6_1}
\end{equation}
where $j^\mu$ is the fermion current of all fermions which carry $\mathrm{SU}_\mathrm{L}(2) \times \mathrm{U}_\mathrm{Y}(1)$ charge.  We observe that the zeroth component of \eqref{eq:O6_1} is the $B+L$ charge density. 

We now consider how an operator of this form can be generated.  Within the Standard Model itself, one can use quark loops and the CP-violating phase of the CKM matrix~\cite{Shaposhnikov:1987tw,Shaposhnikov:1987pf} to generate an effective operator of the form
\begin{equation}
\mathcal{O}_6 = - \dfrac{1}{\Lambda_n^2} \phi^2 \left( g^2 W \tilde{W} - g^{\prime 2} A \tilde{A} \right),
\label{eq:O6_Operator}
\end{equation}
where $W$ and $A$ are the $\mathrm{SU}_\mathrm{L}(2)$ and $\mathrm{U}_\mathrm{Y}(1)$ gauge fields respectively.  This term is small due to the small Yukawa couplings and small CP-violating phase.  

However, a term of the same form can be generated by replacing some or all of the quarks with heavier fermions, which may have larger Yukawa couplings and/or CP-violating phases.  The scale in the denominator may be $T$, due to thermal loops, or the mass scale of these fermions, $M_n$~\cite{Shaposhnikov:1987tw,Shaposhnikov:1987pf,Smit:2004kh,Brauner:2012gu}.  In the latter case, it is important that the fermions not acquire masses through the Higgs mechanism; otherwise, the Higgs VEV dependence in this term cancels out.  Such fermions may acquire soft masses similarly to higgsinos and gauginos in supersymmetric models.

This operator could be generated in a UV-complete model.  As a concrete example, we mention the fully renormalizable Lagrangian
\begin{align}
\mathcal{L}_{hd} &= g \bar{\psi}_{1i} \gamma^\mu \psi_{1i} W_\mu  + g^\prime \bar{\psi}_{1i} \gamma^\mu \psi_{1i} A_\mu + y_{i} e^{i \delta_{i}} \phi \bar{\psi}_{1i} \psi_{2} \nonumber \\
& + M_{ij} \bar{\psi}_{1i} \psi_{1j} + m \bar{\psi}_2 \psi_2 + h.c., 
\label{eq:Fermion_Ex}
\end{align}
where $\psi_{1i}$ are a set of $\mathrm{SU}(2)$ doublets, while $\psi_2$ is a singlet under both $\mathrm{SU}(2)$ and $\mathrm{U}(1)$.  Despite the explicit mass terms, this Lagrangian is invariant under $\mathrm{SU}(2)$ rotations provided that both right and left components of the $\psi_{1i}$ doublets couple vectorially to gauge bosons.  The phases of the $\psi_{1i}$ doublets may be fixed by eliminating the phases in the mass matrix.  Provided that $i \geq 3$, there are more phases $\delta_i$ than can be eliminated by rotating the Higgs field $\phi$ and the singlet $\psi_2$.  Fermionic loops  such as those in~\cite{Shaposhnikov:1987tw,Shaposhnikov:1987pf}, which involve sums of the Yukawa couplings $y_i e^{i \delta_i}$ due to insertions of the Higgs VEV $\left< \phi \right>$ generate an effective operator of the form \eqref{eq:O6_Operator}.  In this case, the scale in the $\mathcal{O}_6$ operator is $\Lambda_n \sim M \sim m$.  

Once an effective operator of the form \eqref{eq:O6_Operator} is generated, it may be transformed into \eqref{eq:O6_1} through the electroweak anomaly equation~\cite{Dine:1990fj}.  However, this is only justified if the electroweak sphalerons are in thermal equilibrium~\cite{Ibe:2015nfa,Daido:2015gqa}.  Otherwise, the operator \eqref{eq:O6_Operator} involves the Chern-Simons number density, which is not changed by Higgs relaxation unless the phase of the Higgs VEV evolves.  At least for slowly evolving Higgs VEVs, the sphaleron transition rate per unit volume at finite temperature is
\begin{equation}
\Gamma_\mathrm{sp} = k \alpha_W^5 T^4 \exp(-v \slash 2 T),
\end{equation}
where the exponential factor accounts for the suppression due to being in the broken phase.  As both the Higgs VEV and the temperature are quickly evolving in the scenario considered here, it may be difficult to arrange for the electroweak sphalerons to be in thermal equilibrium.  However, additional gauge groups which couple to fermions can contribute to the anomaly and generate the requisite term, as discussed in Appendix A in \cite{Pearce:2015nga}.  For our purposes, we simply consider a scenario with operator \eqref{eq:O6_1} without specifying the mechanism by which it is generated.

Returning to equation \eqref{eq:O6_1}, we observe that integrating by parts and dropping an unimportant boundary term gives
\begin{equation}
\mathcal{O}_6 = - \partial_\mu \left( \dfrac{\phi^2}{\Lambda_n^2}\right) j^\mu.
\end{equation}
In the case where $\Lambda_n = M_n$ a constant (for example, the mass scale of a fermionic loop, as outlined around Eq.~\eqref{eq:Fermion_Ex}), this becomes
\begin{equation}
\mathcal{O}_{6,M_n} = - \dfrac{1}{M_n^2} (\partial_\mu \phi^2) j^\mu.
\label{eq:O6_with_mass}
\end{equation}
If thermal loops generate this term instead, then this becomes
\begin{equation}
\mathcal{O}_{6,T} = - \partial_\mu \left( \dfrac{\phi^2}{T^2} \right) j^\mu \approx - \dfrac{1}{T^2} (\partial_\mu \phi^2) j^\mu,
\label{eq:O6_with_T}
\end{equation}
provided that the temperature is slowly varying on the time scales of the Higgs oscillation.  In the IC-1 scenario specifically, the Higgs VEV remains trapped until there is sufficient reheating, which generally ensures that the temperature will be slowly varying during the evolution of the Higgs condensate.  Since the Higgs VEV varies only in time, these equations become
\begin{align}
\mathcal{O}_{6,\Lambda_{n}} &= - \dfrac{1}{\Lambda_n^2} (\partial_0 \phi^2) j_{B+L}^0.
\end{align}
For each fermionic species, its contribution to this term can be combined with its kinetic energy term, $\bar{\psi}(i \slashed{\partial}) \psi$, which is equivalent to the replacement
\begin{equation}
i \partial_0 \rightarrow i \partial_0 - (\partial_0 \phi^2) \slash \Lambda_n^2.
\end{equation}
This effective raises the energy of antiparticles, $E \rightarrow E +  (\partial_0 \phi^2) \slash \Lambda_n^2$, while lowering it for particles, $E \rightarrow E - (\partial_0 \phi^2) \slash \Lambda_n^2$.  This can be interpretted as an external chemical potential; further remarks along these lines are discussed in Appendix~\ref{sec:chem_potential_apndx}.  In the presence of a lepton-number-violating interaction, the system will relax to its equilibrium state, in which the number of particles exceeds the number of antiparticles.  For future reference, we define the effective external chemical potential,
\begin{equation}
E_0 = \dfrac{\partial_0 \phi^2}{\Lambda_n^2}.
\end{equation}

As an effective chemical potential, this operator spontaneously breaks not only $CP$, but in fact, $CPT$~\cite{Cohen:1987vi}.  This operator has been used previously in spontaneous baryogenesis models utilizing gauge~\cite{Dine:1990fj,GarciaBellido:1999sv,GarciaBellido:1999px,GarciaBellido:2003wd,Tranberg:2003gi}  or gravitational~\cite{Davoudiasl:2004gf} interactions.

\section{Lepton Number Violating Processes}
\label{sec:lepton_number_violating}

\begin{figure}
\includegraphics[width=1\columnwidth]{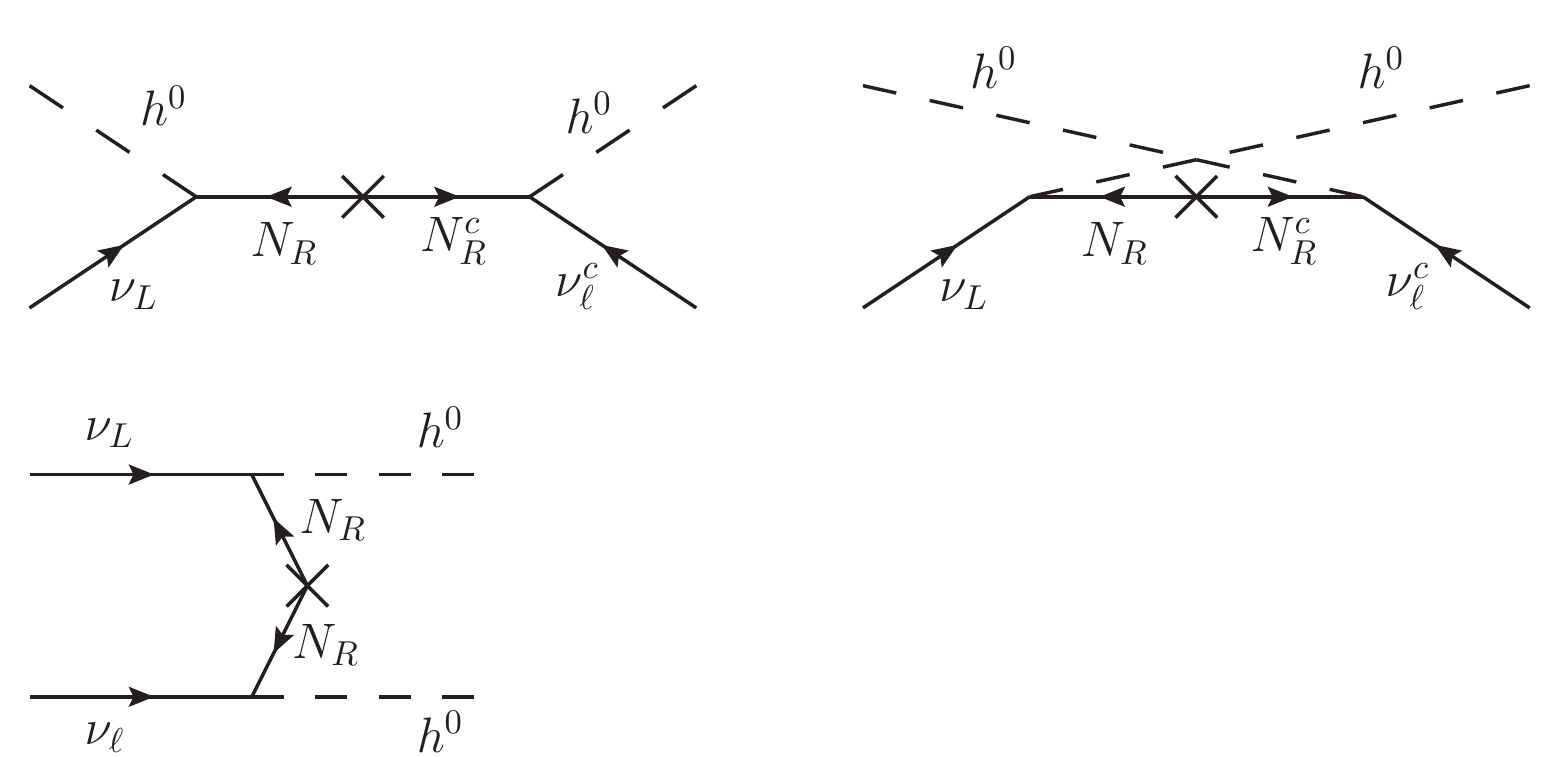}
\caption{Some diagrams that contribute to lepton number violation via exchange of a heavy Majorana neutrino.}
\label{fig:lepton_violation}
\end{figure}

The universe can relax to its equilibrium state with nonzero lepton and baryon number only if there exists some lepton-number or baryon-number violating process.  In order to induce such processes, we consider a minimal extension of the Standard Model with the usual seesaw mass matrix in the neutrino sector~\cite{Yanagid1979,*Yanagida:1980xy,*Gell-Mann1979}.  In theories with a nonzero Majorana mass, the effective lepton number $L$ is the sum of the lepton numbers of the charged leptons and the helicities of the light neutrinos.  This is conserved in the limits $M_R \rightarrow \infty$ and $M_R \rightarrow 0$, but it is not conserved for a finite $M_R$.  Insertions of the Majorana mass induce lepton-number-violating processes such as those shown in Fig.~\ref{fig:lepton_violation}.

We further require that the Majorana mass $M_R$ be significantly greater than both the maximum reheat temperature and the initial mass of the Higgs bosons within the condensate ($m_\mathrm{eff}(\phi_0)$), which suppresses the production of right-handed neutrinos both from thermal production and the decay of the Higgs condensate.  Consequently, the contribution from the typical leptogenesis scenario~\cite{Fukugita:1986hr} is strongly suppressed.

The lepton-number-violating diagrams shown in Fig.~\ref{fig:lepton_violation} necessarily involve the exchange of the heavy right-handed neutrino in order to violate lepton number, and therefore are comparatively suppressed, leading to a naturally small value for the asymmetry.

In order to calculate the thermally averaged cross section $\left< \sigma v \right>$ for these processes in the early universe, we need to know the number densities of neutrinos and Higgs bosons.  These can be produced directly through the decay of the inflaton, or in the thermal plasma through weak interactions, involving weak bosons with masses $m_W \propto \phi(t)$.  Generically these weak interactions may be in or out of equilibrium the plasma created by inflaton decay; however, when $\phi(t) \sim 0$, these interactions will be in equilibrium and equilibrate the distributions of charged and neutral leptons.  To be concrete, we will use a thermal number density of each of these species.  The calculation of the cross section and reaction rate are given in Appendix \ref{sec:cross_section_reaction_rate_apndx}.

We note that $y^2 \slash M_R$ is set by the mass scale of the left-handed neutrinos, such that
\begin{align}
\dfrac{y^2 v_{\mathrm{EW}}^{2}}{2 M_R} & = 0.1 \; \mathrm{eV}.
\label{eq:neutrino_mass}
\end{align}
The cross section found in Appendix \ref{sec:cross_section_reaction_rate_apndx} is to a good approximation a function of $y^2 \slash M_R$ only.  There is a resonance in the $s$-channel contribution to the cross section; however, we found numerically that this resonance does not change the result appreciably.  This is not unexpected as the energy scale, which is set by the temperature, remains significantly below the right-handed Majorana mass scale at all times.

As we will discuss below, sphaleron processes later convert this lepton charge asymmetry into a baryon asymmetry, as in the typical leptogenesis scenario~\cite{Fukugita:1986hr}.

\section{Boltzmann Transport Equation}
\label{sec:boltzmann_equation_ahhhh!}

The reactions discussed in Section \ref{sec:lepton_number_violating} are generally not sufficient to establish equilibrium, due to the suppression from the large Majorana mass.  (Recall that we have assumed $\phi_0 \ll M_R$ and $T_\mathrm{max} \ll M_R$ in order to suppress the typical leptogenesis mechanism.)  The relaxation of the system towards equilibrium can be described by a system of Boltzmann equations, based on detailed balance.  The rate of change in the neutrino number density is~\cite{Giudice:2003jh}
\begin{widetext}
\begin{align}
 & \dot{n}_{\nu_{L}}+3Hn_{\nu_{L}}=-\sum_{\ell=e,\mu,\tau}\left[\dfrac{n_{\nu_{L}}n_{h^{0}}}{n_{\nu_{L}}^{eq}n_{h^{0}}^{eq}}\gamma^{eq}(\nu_{L}h^{0}\rightarrow\bar{\nu}_{\ell}h^{0})-\dfrac{n_{\bar{\nu}_{\ell}}n_{h^{0}}}{n_{\bar{\nu}_{\ell}}^{eq}n_{h^{0}}^{eq}}\gamma^{eq}(\bar{\nu}_{\ell}h^{0}\rightarrow\nu_{L}h^{0})\right.\nonumber \\
 & \left.+\dfrac{n_{\nu_{L}}n_{\nu_{\ell}}}{n_{\nu_{L}}^{eq}n_{\nu_{\ell}}^{eq}}\gamma^{eq}(\nu_{L}\nu_{\ell}\rightarrow h^{0}h^{0})-\dfrac{n_{h^{0}}^{2}}{n_{h^{0}}^{eq\,2}}\gamma^{eq}(h^{0}h^{0}\rightarrow\nu_{L}\nu_{\ell})+\dfrac{n_{\nu_{L}}n_{\bar{\nu}_{\ell}}}{n_{\nu_{L}}^{eq}n_{\bar{\nu}_{\ell}}^{eq}}\gamma^{eq}(\nu_{L}\bar{\nu}_{\ell}\rightarrow h^{0}h^{0})-\dfrac{n_{h^{0}}^{2}}{n_{h^{0}}^{eq\,2}}\gamma^{eq}(h^{0}h^{0}\rightarrow\nu_{L}\bar{\nu}_{\ell})\right],
\end{align}
where $\gamma^{eq}(A \rightarrow B)$ is the equilibrium spacetime rate for the process $A \rightarrow B$.  We will assume that interactions are sufficiently fast that the Higgs bosons have their equilibrium density, and in equilibrium, the rate for the process $A \rightarrow B$ is equal to the rate of $B \rightarrow A$.  Therefore this simplifies to
\begin{align}
\dot{n}_{\nu_{L}}+3Hn_{\nu_{L}} & =-\sum_{\ell=e,\mu,\tau}\left[\left(\dfrac{n_{\nu_{L}}}{n_{\nu_{L}}^{eq}}-\dfrac{n_{\bar{\nu}_{\ell}}}{n_{\bar{\nu}_{\ell}}^{eq}}\right)\gamma^{eq}(\nu_{L}h^{0}\leftrightarrow\bar{\nu}_{\ell}h^{0})+\left(\dfrac{n_{\nu_{L}}n_{\nu_{\ell}}}{n_{\nu_{L}}^{eq}n_{\nu_{\ell}}^{eq}}-1\right)\gamma^{eq}(\nu_{L}\nu_{\ell}\leftrightarrow h^{0}h^{0})\right.\nonumber \\
 & \quad\left.+\left(\dfrac{n_{\nu_{L}}n_{\bar{\nu}_{\ell}}}{n_{\nu_{L}}^{eq}n_{\bar{\nu}_{\ell}}^{eq}}-1\right)\gamma^{eq}(h^{0}h^{0}\leftrightarrow\nu_{L}\bar{\nu}_{\ell})\right],\label{eq:n_nu_L}
\end{align}
while for antineutrinos we find the similar equation
\begin{align}
\dot{n}_{\bar{\nu}_L} + 3 H n_{\bar{\nu}_L} &= -\sum_{\ell = e,\mu,\tau} \left[\left( \dfrac{n_{\bar{\nu}_L} }{n_{\bar{\nu}_L}^{eq} } -  \dfrac{n_{\nu_\ell}}{n_{\nu_\ell}^{eq} } \right) \gamma^{eq}(\bar{\nu}_L h^0 \leftrightarrow \nu_\ell h^0) + \left( \dfrac{n_{\bar{\nu}_L} n_{\bar{\nu}_\ell}}{n_{\bar{\nu}_L}^{eq} n_{\bar{\nu}_\ell}^{eq}} - 1 \right) \gamma^{eq}(\bar{\nu}_L \bar{\nu}_\ell \leftrightarrow h^0 h^0) \right. \nonumber \\
& \quad\left.  + \left( \dfrac{n_{\bar{\nu}_L} n_{\nu_\ell}}{n_{\bar{\nu}_L}^{eq} n_{\nu_\ell}^{eq}}  - 1 \right) \gamma^{eq}(h^0 h^0 \leftrightarrow \bar{\nu}_L \nu_\ell) \right].\label{eq:n_nu_bar_L}
\end{align}
Since we are interested in the order of magnitude of the final asymmetry,
we simplify to the case in which there is only a single neutrino species.
Subtracting Eq.\ \eqref{eq:n_nu_L} from Eq.\ \eqref{eq:n_nu_bar_L}
gives a Boltzmann-type equation for the difference $n_{L}=n_{\nu_{L}}-n_{\bar{\nu}_{L}}$,
\begin{align}
\dot{n}_{L}+3Hn_{L} & =-2\left(\dfrac{n_{\nu_{L}}}{n_{\nu_{L}}^{eq}}-\dfrac{n_{\bar{\nu}_{L}}}{n_{\bar{\nu}_{L}}^{eq}}\right)\gamma^{eq}(\nu_{L}h^{0}\leftrightarrow\bar{\nu}_{L}h^{0})-\left(\dfrac{n_{\nu_{L}}^{2}}{n_{\nu_{L}}^{eq\,2}}-1\right)\gamma^{eq}(\nu_{L}\nu_{L}\leftrightarrow h^{0}h^{0})\nonumber \\
 & \quad+\left(\dfrac{n_{\bar{\nu}_{L}}^{2}}{n_{\bar{\nu}_{L}}^{eq\,2}}-1\right)\gamma^{eq}(\bar{\nu}_{L}\bar{\nu}_{L}\leftrightarrow h^{0}h^{0}).
\end{align}
\end{widetext}
The rates $\gamma^{eq}(A \leftrightarrow B)$ refer to the process $A \leftrightarrow B$ in equilibrium, but in the presence of the $\mathcal{O}_6$ operator, which alters the energy of particles and antiparticles.  Consequently, these reaction rates are not generally equal to the rates one would find in the absence of the $\mathcal{O}_6$ operator; however, the difference appears at a higher order in $E_0 \slash T$~\cite{Ibe:2015nfa} and so we will neglect it.  This has the consequence that the rates for $h^0 h^0 \leftrightarrow \nu_L \nu_L$ and $h^0 h^0 \leftrightarrow \bar{\nu}_L \bar{\nu}_L$ are equal.  We will use the subscript 0 to denote reaction rates calculated without the $\mathcal{O}_6$ operator.

We next substitute $n_{\nu_{L}}^{eq}=e^{E_{0}\slash T}n_{0}^{eq}$ and $n_{\bar{\nu}_{L}}^{eq}=e^{-E_{0}\slash T}n_{0}^{eq}$, where $n^{eq}_0 = T^3 \slash \pi^2$ is the equilibrium number of left-handed neutrinos (or antineutrinos), when $E_0 =0$.  Expanding the resulting equation to lowest order in $E_0 \slash T$ gives
\begin{multline}
\dot{n}_{L}+3Hn_{L}=-\dfrac{2}{n_{0}^{eq}}\left(n_{L}-\dfrac{E_{0}}{T}n_{L}^{\mathrm{tot}}\right)\gamma_{0}^{eq}(\bar{\nu}_{L}h^{0}\leftrightarrow\nu_{L}h^{0})\\
-\dfrac{1}{n_{0}^{eq\,2}}\left(n_{L}^{\mathrm{tot}}n_{L}-\dfrac{E_{0}}{T}n_{L}^{\mathrm{tot}\,2}\right)\gamma_{0}^{eq}(\nu_{L}\nu_{L}\leftrightarrow h^{0}h^{0}),
\end{multline}
where we have introduced the notation $n_{L}^{\mathrm{tot}}=n_{\nu_{L}}+n_{\bar{\nu}_{L}}$, and we have dropped terms quadratic in the asymmetry (e.g., $n_{L}^{2}$).  Approximating $n_{L}^{\mathrm{tot}}\approx2n_{0}^{eq}$, the equation becomes 
\begin{align}
\dot{n}_{L}+3Hn_{L} & =-\dfrac{2}{n_{0}^{eq}}\left(n_{L}-\dfrac{2E_{0}}{T}n_{0}^{eq}\right)\left[\gamma_{0}^{eq}(\bar{\nu}_{L}h^{0}\leftrightarrow\nu_{L}h^{0})\right.\nonumber \\
 & \quad\left.+\gamma_{0}^{eq}(\nu_{L}\nu_{L}\leftrightarrow h^{0}h^{0})\right].
\label{eq:nL}
\end{align}
The reaction rates are calculated in Appendix~\ref{sec:cross_section_reaction_rate_apndx}.  From this equation, we observe that the equilibrium asymmetry is
\begin{equation}
n_{L,eq} = \dfrac{2 E_0}{T} n_0^{eq} = \dfrac{2 T^2}{\pi^2} \dfrac{\partial_0 \phi^2}{\Lambda_n^2}.
\end{equation}

\begin{figure}
\includegraphics[width=1\columnwidth]{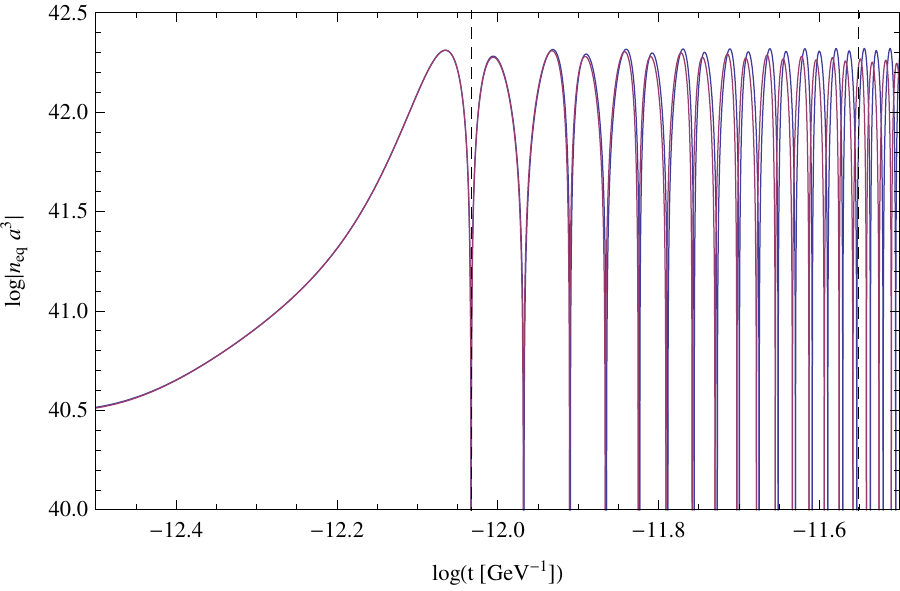} \\
\includegraphics[width=1\columnwidth]{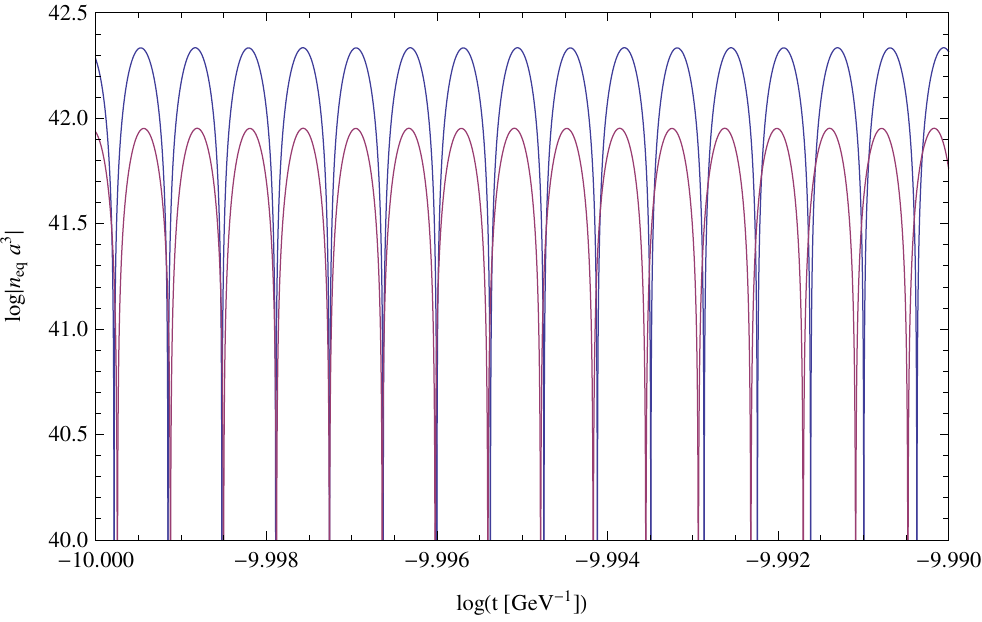}
\protect\caption{The comoving density of equilibrium lepton asymmetry for IC-1, with the parameters $\Lambda_{I}=10^{15}\;\mathrm{GeV}$ and $\Gamma_{I}=10^{9}\;\mathrm{GeV}$. Purple (Blue) line corresponds to the result with (without) the thermalization through the top quark.  The top diagram corresponds to times before maximum reheating, whereas the bottom diagram corresponds to times after maximum reheating.}
\label{equilibrium lepton asymmetry for thermalization}
\end{figure}

During subsequent oscillations of the Higgs VEV, the chemical potential changes sign.  However, due to the large suppression in the cross section, significant washout can be avoided if the Higgs oscillation amplitude decreases rapidly.  This is in contrast to Ref.~\cite{GarciaBellido:1999px}, in which washout was avoided by using coherent oscillations of the inflaton field to modify the sphaleron transition rate.

We note that this is modified by the decay of the Higgs condensate; however, as discussed above, the Higgs condensate does not typically thermalize until after reheating.  Fig.~\ref{equilibrium lepton asymmetry for thermalization} demonstrates the effect of thermalization on the equilibrium density.  However, since the lepton asymmetry will be generated primarily during the first oscillation of the Higgs VEV, the effect of the thermalization of the Higgs condensate is negligible.

\section{Resulting Asymmetry}
\label{sec:asymmetry_produced}

In this section, we consider the lepton asymmetry produced by these Higgs-neutrino interactions during the relaxation of the Higgs VEV, as outlined above.  We present four numeric examples, covering both IC-1 and IC-2, along with the scale of the $\mathcal{O}_6$ operator $\Lambda_n$ set to the temperature $T$ (motivated by thermal loops) and a constant $M_n$ (motivated by loops of heavy fermions).  This expands the analysis of~\cite{Kusenko:2014lra}, which only considered two such scenarios.  In all scenarios, we use the improved Boltzmann equation \eqref{eq:nL}, with the cross sections calculated in Appendix \ref{sec:cross_section_reaction_rate_apndx}, and the improved calculation of the Higgs condensate equation of motion.   We show the time-evolution of the lepton asymmetry in all four scenarios; subsequently, we present an analysis of the parameter space in which a sufficiently large late-time lepton asymmetry can be generated.  

An analytic approximation for the asymmetry calculated here numerically can be found in~\cite{Kusenko:2014lra}, which we summarize here.  The Boltzmann equation \eqref{eq:nL} can be analyzed in two regimes: during the relaxation of the Higgs vacuum expectation value, during which $\partial_t \phi^2$ is significant and $E_0(t) \neq 0$, and the subsequent cooling of the universe, during which $E_0 = 0$.  The reactions shown in Fig.~\ref{fig:lepton_violation} are typically out of thermal equilibrium by the end of Higgs relaxation, due to the exchange of the heavy right-handed neutrino, which suppresses washout.  One may approximate the potential as $V \sim \lambda \phi^4$, with an effective running coupling $\lambda$ as in~\cite{Degrassi:2012ry}.  The final asymmetry  $\eta = n_L/(2 \pi^2 g_*T^3 \slash 45)$ is approximately
\begin{align}
\eta & = \dfrac{45}{2\pi^2} \frac{\sqrt{\lambda}\phi_{0}^{3} \Lambda_I}{M_{n}^{2}T_R^2}\ 
t_{\textrm{rlx}}^2 \Gamma_I^2 \times \min \left \{1,  T_{\textrm{rlx}}^{3} t_{\textrm{rlx}} \sigma_{R} \right\} \nonumber
\\ 
& \quad\times \exp\left[-\left( \frac{24 + 3\sqrt{15}}{\sqrt{ 3 g_* \pi^7}} \right) \sigma_R M_{\mathrm{Pl}} T_R\right],
\label{eq:eta_analytical}
\end{align}
where $t_\mathrm{rlx}$ and $T_\mathrm{rlx}$ as the time and temperature at the end of Higgs relaxation, and $\sigma_R \approx 10^{-31} \; \mathrm{GeV}^{-2}$ approximates the the cross section given by equation \eqref{eq:cross_section}.  This estimate includes the dilution due to entropy production during the ongoing reheating process.

\subsection{Four Numerical Examples}

\begin{figure}
\includegraphics[scale=.65]{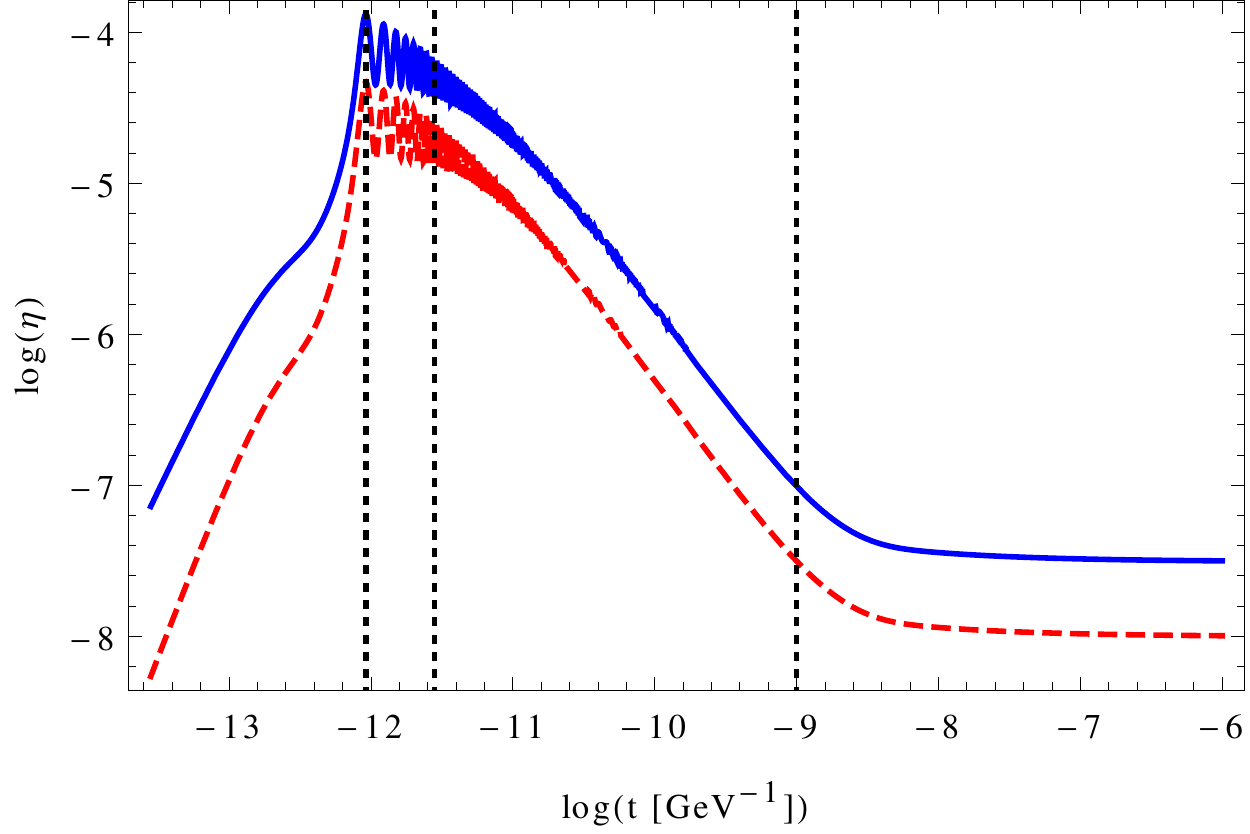}
\caption{Plot of the resulting asymmetry for IC-1, for $\Lambda_n = T$ (blue, solid) and $\Lambda_n = M_n = 10^{14}$ GeV (red, dashed).  Both scenarios have $\Lambda_I = 10^{15}$ GeV, and $\Gamma_I = 10^9$ GeV.  The vertical lines designate the first time the Higgs VEV crosses zero, time of maximum reheating, and the beginning of the radiation dominated era, from left to right.  $t=0$ corresponds to the beginning of inflaton oscillations.}
\label{fig:IC_1_plots}
\end{figure}

In this subsection, we present the lepton asymmetry as a function of time for the four scenarios mentioned above.  First, we consider two scenarios for IC-1 in Fig.~\ref{fig:IC_1_plots}, one with  $\Lambda_n = T$ (blue, solid) and one with $\Lambda_n = M_n = 10^{14}$ (red, dashed) for the relevant scales in the $\mathcal{O}_6$ operator.  Both scenarios have a maximum temperature of $ 6 \times 10^{13}$~GeV, since they share the inflationary parameters $\Lambda_I = 10^{15}$ GeV and $\Gamma_I = 10^9$ GeV.  As in Fig.~\ref{fig:Higgs_Evolution_IC1}, the initial Higgs VEV is $10^{15}$~GeV in both cases, which is set by the location of the second minimum in the Higgs potential.  Although the asymmetry $\eta$ oscillates during the first few oscillations of the Higgs VEV, it relatively quickly settles into a steady state, and approaches a constant value around the beginning of the radiation dominated era.  Note that the Higgs field begins to oscillate before the time of maximum reheating.  

As mentioned above, the cross section depends primarily on  $y^2 \slash M_R$ which is fixed by the light neutrino masses.  As mentioned above, we require $T \ll M_R$ in order to suppress the thermal production of right-handed neutrinos; we found that it was sufficient to set $M_R = 9 \times 10^{15}$~GeV, which results in $y \sim 1.7$ using equation \eqref{eq:neutrino_mass}.  (This gives $y^2 \slash 4 \pi \sim 0.2$, within the perturbative regime.)

The late time asymptotic asymmetry is $\eta \sim 10^{-7}$ for $\Lambda_n = T$ and $\eta \sim 10^{-8}$ for $\Lambda_n = M_n = 10^{14} \; \mathrm{GeV}$; this is expected as the temperature is lower than $M_n$.  We discuss the variation of the final asymmetry over parameter space below.

\begin{figure}
\includegraphics[scale=.65]{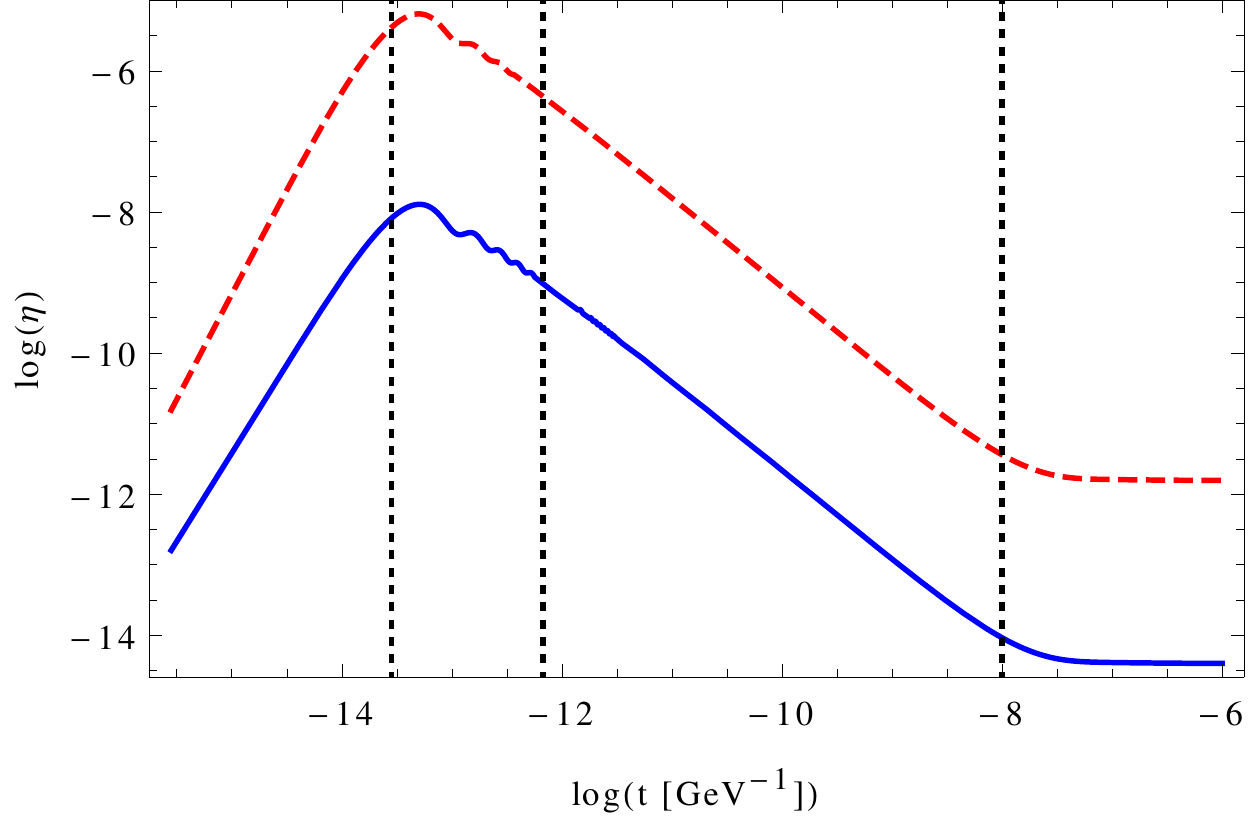}
\caption{Plot of the resulting asymmetry for IC-2, for $\Lambda_n = T$ (blue, solid) and $\Lambda_n = M_n = 5 \times 10^{12}$~GeV (red, dashed).  Both scenarios have $\Lambda_I = 10^{16}$ GeV and $\Gamma_I = 10^8$~GeV.  From left to right, the dotted lines correspond to the time of maximum reheating, the first time the Higgs VEV crosses zero, and the beginning of the radiation dominated era.}
\label{fig:IC_2_plots}
\end{figure}

First, however, we present similar results for the IC-2 scenario, again for the two cases $\Lambda_n=T$ (blue, solid) and $\Lambda_n = M_n = 5 \times 10^{12}$~GeV (red, dashed).  Both plots have the inflationary parameters $\Lambda_I = 10^{16}$ GeV and $\Gamma_I = 10^8$~GeV, which results in a maximum temperature of $10^{14} \; \mathrm{GeV}$ during reheating.  We again take $N_\mathrm{last} = 8$ to determine the Higgs VEV at the end of inflation; this results in $\phi_0 = 10^{13}$~GeV for the Higgs VEV as the start of Higgs relaxation.  (We emphasize that this choice, with $M_n < \phi_0$ and $M_n < T$, raises questions regarding the use of effective field theory, which we address below.)

In order to suppress the thermal production of right-handed neutrinos, we have taken $M_R = 10 T_\mathrm{max} = 10^{15} \; \mathrm{GeV}$; in order to produce left-handed neutrino masses on the scale of $0.1 \; \mathrm{eV}$, the neutrino Yukawa coupling must be $~1.9$.  (This gives $y^2 \slash 4 \pi \approx 0.3$.)

The final asymmetries here are of order $10^{-14}$ (for $\Lambda_n = T$) and $10^{-12}$ (for $\Lambda_n = M_n = 5 \times 10^{12}$~GeV).  As $M_n$ is generally smaller than the temperature, it is not surprising that this results in a larger asymmetry.  These values are insufficient to account for the observed matter-antimatter asymmetry; this motivates a search of the available parameter space.

\subsection{Parameter Space}

\begin{figure}
\includegraphics[width=1\columnwidth]{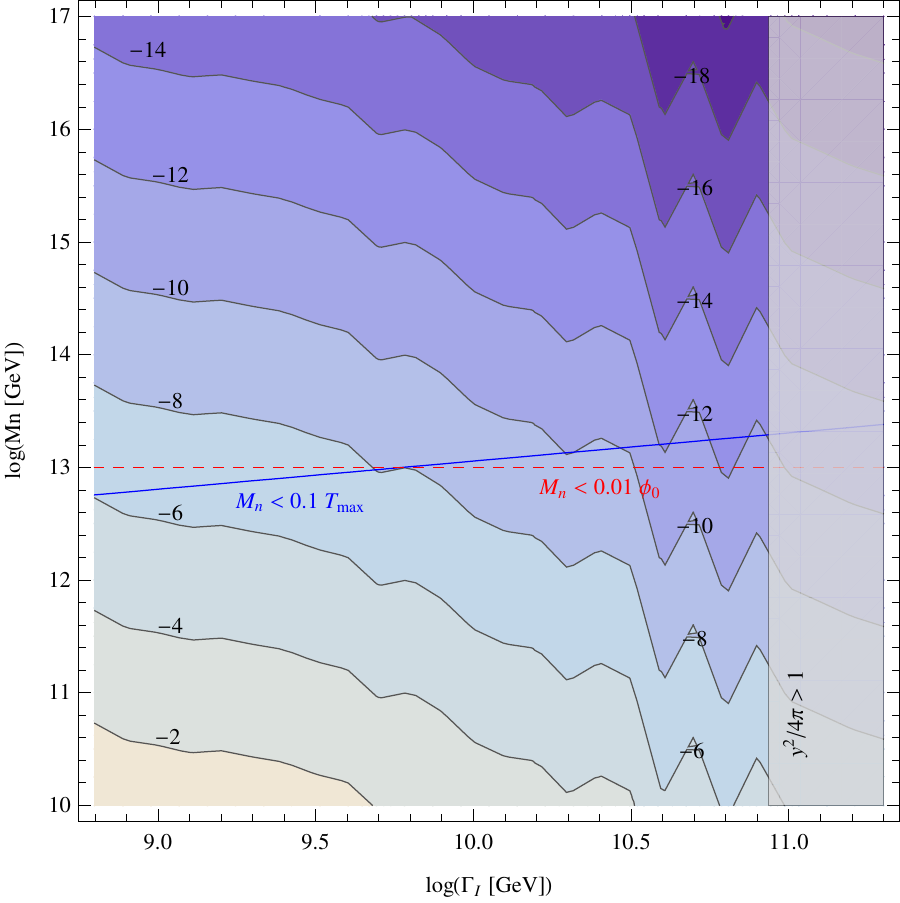}
\protect\caption{The resulting asymmetry ($\log\left|\eta\right|$) at the end of reheating
for IC-1, for $\Lambda_{n}=M_{n}$, with $\Lambda_{I}=10^{15}$ GeV.
\label{fig:Parameter space IC1 Mn}}
\end{figure}

\begin{figure}
\includegraphics[width=1\columnwidth]{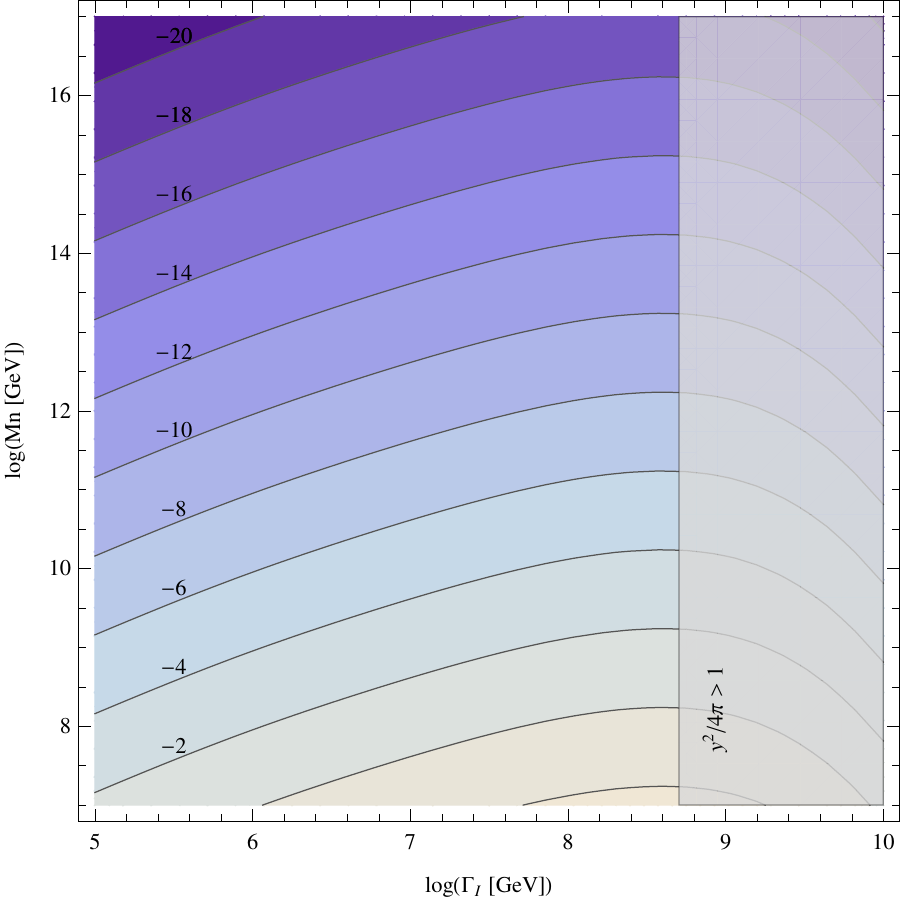}
\protect\caption{The resulting asymmetry ($\log\left|\eta\right|$) at the end of reheating for IC-2, for $\Lambda_{n}=M_{n}$ with $\Lambda_{I}=5\times10^{16}$ GeV, which gives $\phi_{0}=2.7\times10^{14}$ GeV.\label{fig:Parameter space IC2 Mn}}
\end{figure}

\begin{figure}
\includegraphics[width=1\columnwidth]{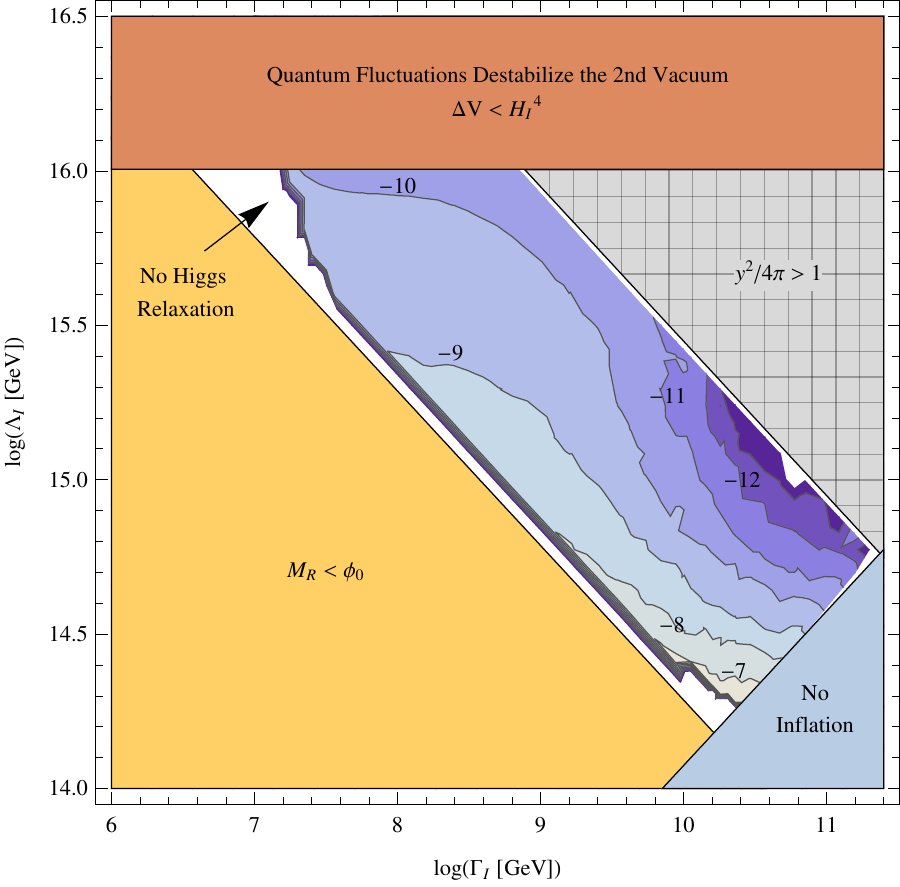}
\protect\caption{The resulting asymmetry ($\log\left|\eta\right|$) at the end of reheating for IC-1, for $\Lambda_{n}=T$.\label{fig:Parameter space IC1 T}}
\end{figure}

\begin{figure}
\includegraphics[width=1\columnwidth]{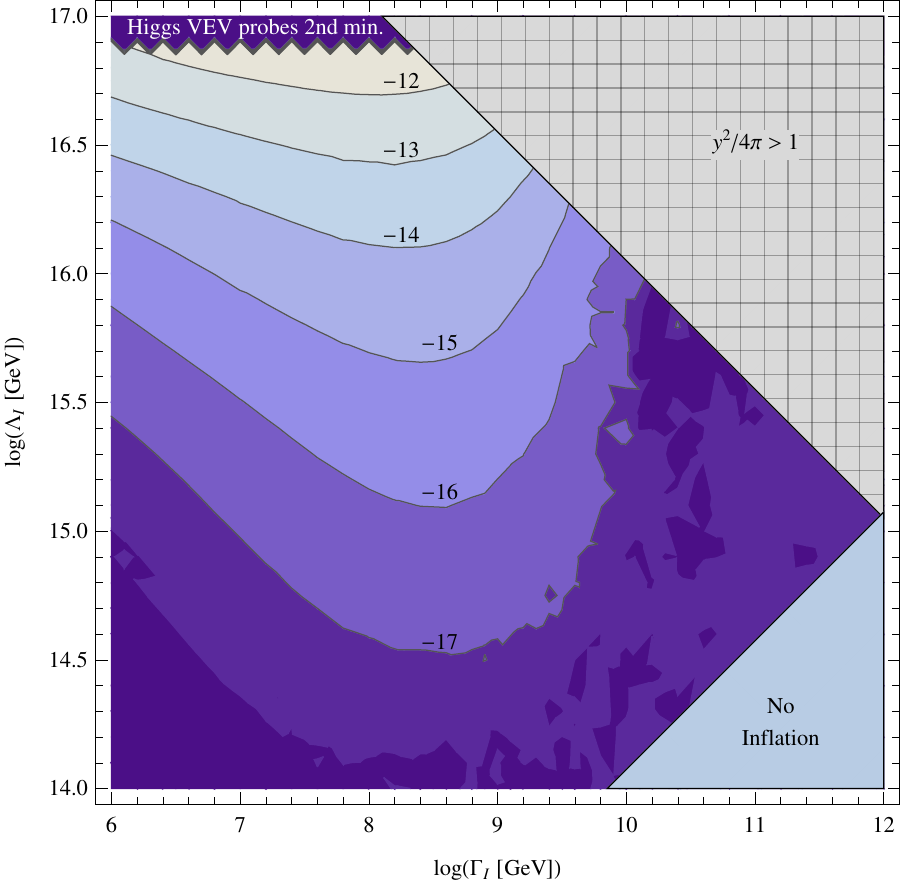}
\protect\caption{The resulting asymmetry ($\log\left|\eta\right|$) at the end of reheating for IC-2, for $\Lambda_{n}=T$. 
\label{fig:Parameter space IC2 T}}
\end{figure}

In two of the four scenarios above, the resulting lepton asymmetry is $\mathcal{O}(10^{-8})$ or larger, which is sufficient to explain the observed baryon asymmetry.  However, it is interesting to explore the resulting asymmetry as a function of parameter space; results are shown in Figures \ref{fig:Parameter space IC1 Mn}, \ref{fig:Parameter space IC2 Mn}, \ref{fig:Parameter space IC1 T}, and \ref{fig:Parameter space IC2 T}.  

As above, we handle the initial conditions with the operator and scale given in \ref{subsec:IC1} for the IC-1 plots, and as discussed in \ref{subsec:IC2} with $N_\mathrm{last} = 8$ for the IC-2 plots.  We emphasize again that the resulting asymmetry is sensitive to $y^2 \slash M_R$, which is set by the left-handed neutrino mass scale, and not on the specific value of $M_R$.  However, to suppress thermal production of right-handed neutrinos, we have chosen $M_R = 10 T_\mathrm{max}$ (for IC-1) and $M_R = 20 T_\mathrm{max}$ (for IC-2).  We have then set the neutrino Yukawa coupling $y$ by the scale of the left-handed neutrino masses (equation \eqref{eq:neutrino_mass}).  We have noted in gray the regions in which the perturbativity condition $y^2 \slash 4\pi < 1$ fails.

For the IC-1 plots, the post-inflationary Higgs VEV $\phi_0$ is determined entirely by the operator which lifts the second minimum to generate the quasistable vacuum; for the operator and scale in \ref{subsec:IC1}, the Higgs VEV relaxes from $\phi_0 = 10^{15} \; \mathrm{GeV}$.  For IC-2, $\phi_0$ is determined by the Hubble parameter during inflation, which is in turn fixed by the energy density in the inflation field (see equation \eqref{eq:v0_IC2}).  

First, we remark on some general features.  The asymmetries generated in the IC-2 scenario are smaller than those generated in the IC-1 scenario.  This is because in IC-1, the Higgs VEV does not evolve until the temperature is sufficiently large to destabilize the false vacuum; consequently, the initial evolution of the VEV to zero occurs at higher temperatures.  (Compare the vertical lines significantly the first Higgs VEV crossing and maximum reheating in Figures \ref{fig:IC_1_plots} and \ref{fig:IC_2_plots}.)  As a result of the higher temperature, the system is driven towards equilibrium at a faster rate (through the Boltzmann equation \eqref{eq:nL}); furthermore, in the $\Lambda_n = T$ scenario, the larger temperature also means that the equilibrium charge density is larger.

Figures \ref{fig:Parameter space IC1 Mn} and \ref{fig:Parameter space IC2 Mn} show the lepton asymmetry $\eta$ as a function of parameter space, in the case in which the scale of the $\mathcal{O}_6$ operator is a constant $M_n$.  To reach comparable asymmetries in the IC-2 scenario, we must decrease the scale $M_n$ significantly, such that throughout this plot, $M_n < \phi_0$ and $M_n < T_\mathrm{max}$.  In the IC-1 plot, these conditions fail below the red dashed line and blue solid line respectively.  In these regions, the use of effective field theory in generating the operator \eqref{eq:O6_1} is questionable.  An ultraviolet completion of the model is necessary to obtain a reliable description of the dynamics in the regime where the temperature exceeds the scale $M_{n}$.  We leave such a completion, which would also elucidate the nature of the new physics leading to the appearance of the $\mathcal{O}_{6}$ operator, for a future work.

We focus on the region of Fig.~\ref{fig:Parameter space IC1 Mn} for which the asymmetry $\eta$ is larger than $10^{-10}$ and $M_n > 0.1 T_\mathrm{max}$.  We see that this favors smaller values of $\Gamma_I$.  However, for a given $\Lambda_I$, there is a minimum $\Gamma_I$, for which the maximum temperature is insufficient to destabilize the second vacuum.  For the parameters considered here ($\Lambda_{I} = 10^{15} \; \mathrm{GeV}$ and the lift operator given in \ref{subsec:IC1}), this occurs for $\Gamma_{I} = 6.3\times10^{8}$. 

Next, we consider the case in which the scale of the $\mathcal{O}_6$ operator is set by the temperature, in Figures \ref{fig:Parameter space IC1 T} and \ref{fig:Parameter space IC2 T}.  
This parameter space has one fewer parameters, and so we allow $\Lambda_I$ to also vary, which changes the Hubble parameter during inflation.  For IC-2, increasing $H_I$ results in a larger value of $\phi_0$, as described by \eqref{eq:v0_IC2}, which increases the resulting asymmetry.   
This also increases the temperature scale, resulting in a larger asymmetry, as is evident in both figures.  (We also note that for IC-1, we must take care that quantum fluctuations during inflation do not destabilize the second vacuum; this is shown in orange in Fig.~\ref{fig:Parameter space IC1 T}.) 

As mentioned above, if the reheat temperature is sufficiently small, thermal corrections are unable to destabilize the second vacuum, and therefore this is no relaxation of the Higgs VEV.  This region is denoted in white in Fig.~\ref{fig:Parameter space IC1 T}.  Furthermore, in the region in which $M_R < \phi_0$, right-handed neutrinos can be copiously produced by the decay of the Higgs condensate, which is not desirable (as concerns the lepton asymmetry production scenario presented here); this region is denoted in yellow.  Furthermore, if $\Lambda_{I}$ is too small, there is insufficient inflation to account for the observed flatness and uniformity of the universe; this region is shown in blue on both figures.  

In IC-2, there is a further concern that the Higgs VEV can probe the second, deeper minimum at large VEVs.  This may not be a phenomenological problem~\cite{Kearney:2015vba}, but would require a refinement of the analysis presented here.  (Alternatively, $N_\mathrm{last}$ could be decreased, such that $\phi_0$ remains below the instability scale.)  This region is shown in purple in Fig.~\ref{fig:Parameter space IC2 T}.

We see that for IC-1, it is possible to find parameter space in which a sufficiently large asymmetry is generated, but this is not possible for IC-2.  For IC-1, smaller $\Gamma_I$ values are favored (and consequently, slower reheating), as for constant $\Lambda_n$.

\subsection{Converting the Lepton Asymmetry Into a Baryon Asymmetry}

Thus far, we have analyzed the production of an excess of leptons over antileptons; here we discuss how this is converted into a baryon asymmetry.  First, though as the universe continues to cool, the Standard Model degrees of freedom go out of thermal equilibrium; the resulting entropy production reduces the asymmetry by about two orders of magnitude.

This process has produced a net density of $(B-L)$ charge, which is unchanged once these processes are negligible.  However, the $(B+L)$ $U(1)$ symmetry is anomalous, and electroweak sphalerons will redistribute the excess between leptons and baryons as in standard leptogenesis~\cite{Fukugita:1986hr}, at a rate per unit volume
\begin{equation}
\Gamma_{\mathrm{sp}}\sim(\alpha_{W}T)^{4}\exp\left[-g_{W}\phi(t)/T\right].
\end{equation}
At small vacuum expectation values, the $B$ and $L$ densities approach their equilibrium values, $n_{B}=(28/79)n_{B-L}$.  This produces a baryon asymmetry of about the same order of magnitude as the lepton asymmetry found above.  Consequently, the regions of parameter space that generate $\eta \sim 10^{-8}$ in the analysis above give a final baryon asymmetry matching the observed value of $\mathcal{O}(10^{-10})$.

\section{Conclusions}

In this paper, we have extended the analysis of Ref.~\cite{Kusenko:2014lra}, which introduced a novel leptogenesis possibility in which the lepton asymmetry is as a consequence of an effective chemical potential induced by the post-inflationary relaxation of the Higgs field.  Although right-handed neutrinos participate in the lepton-number-violating interactions as a mediator, this is different from the typical leptogenesis scenario in which the asymmetry is produced via the decay of right-handed neutrinos.  Even though the heavy right-handed neutrino suppresses the cross section which produces the asymmetry, we have shown parameters for which a sufficiently large the asymmetry is generated.  

We have analyzed the evolution of the Higgs condensate in detail, including both non-perturbative and perturbative decay.  We have derived the relevant Boltzmann equation which governs lepton number, and we have replaced the order-of-magnitude estimate with a tree-level scattering cross section between Higgs bosons and neutrinos in the thermal plasma.  Furthermore, we have considered the evolution of the lepton asymmetry for four combinations of producing the large Higgs VEV during inflation (IC-1 and IC-2) and the scale of the $\mathcal{O}_6$ operator (a fermion mass scale $M_n$ and the temperature $T$); we then presented an analysis of the asymmetry as a function of parameter space.  We demonstrated regions which produces a baryonic asymmetry that meets or exceeds observational limits.

\section*{Acknowledgements}

The authors would like to thank K.~Harigaya, M.~Ibe, M.~Kawasaki, M.~Peloso, K.~Schmitz, F.~Takahashi, and T.T.~Yanagida for helpful discussions.  The work of A.K. was  supported by the U.\,S.\ Department of Energy Grant DE-SC0009937, as well as by the World Premier International Research Center Initiative (WPI), MEXT, Japan.

\appendix

\section{Interpretting the $\mathcal{O}_6$ Operator as an External Chemical Potential}
\label{sec:chem_potential_apndx}

In section \ref{sec:chemical_potential}, we remarked that the $\mathcal{O}_6$ operator in equation \eqref{eq:O6_Operator} acts like an external chemical potential.  In this appendix, we explain why this is so and how this leads to a number density asymmetry in chemical equilibrium.

This $\mathcal{O}_6$ operator induces a term proportional to $(\partial_0 \phi^2) \slash \Lambda_n^2 j_{B+L}^0$ in the Lagrangian.  If $\phi$ is treated as an external field (which we discuss further below), then this produces a term of the form $- (\partial_0 \phi^2) \slash \Lambda_n^2 j_{B+L}^0$ in the Hamiltonian, which has the appropriate form $- \mu_\mathrm{eff} j_{B+L}^0$.

A term similar to this, using the phase of the Higgs VEV is frequently used in spontaneous baryogenesis scenarios, in which the phase of the Higgs VEV is used instead of its magnitude (e.g., \cite{Cohen:1990it}),
\begin{equation}
\mathcal{O}_6^\prime = (\partial_t \theta) j_{B+L}^0.
\end{equation}
However, in such scenarios, the asymmetry is produced via the decay of the Higgs condensate, and therefore, it is not appropriate to treat $\theta$ as an external degree of freedom.  When the Hamiltonian is determined using
\begin{equation}
\mathcal{H} = \sum_i \dfrac{\partial \mathcal{L}}{\partial \dot{\phi}_i } \dot{\phi}_i - \mathcal{L},
\end{equation}
there is no contribution from $\mathcal{O}_6^\prime$.  Although an asymmetry may be produced in such cases~\cite{Dolgov:1994zq,Dolgov:1996qq,Dolgov:1997qr}, it is not appropriate to interpret $\dot{\theta}$ as a chemical potential.

In the scenario we consider in this work, the time scale for the reactions which maintain the thermal distribution of the plasma is smaller than that of the evolution of the Higgs VEV. Therefore, for purposes of asymmetry generation, it is reasonable to consider the Higgs VEV as a background field, in which case it is appropriate to consider this as a chemical-potential-like term~\cite{Dolgov:1997qr}, as we explain below.


The $\mathcal{O}_6$ operator shifts $i \partial_0 \rightarrow i \partial_0 - (\partial_0 \phi^2) \slash \Lambda_n^2$ in the Lagrangian.  Consequently, the asymptotically free eigenfunctions are $\sim \exp(\mp i (E \mp (\partial_0 \phi^2) \slash \Lambda_n^2) t)$, which justifies our comment that this is equivalent to decreasing the energy of particles by $E_0 = (\partial_0 \phi^2)\slash\Lambda_{n}^{2}$ and increasing the energy of antiparticles by the same amount.

If we use the ideal gas approximation, then the phase space densities are
\begin{align}
f_p &= \exp(-(E-E_0-\mu_p)\slash T) \nonumber \\
f_{\bar{p}} &= \exp(-(E+E_0-\mu_{\bar{p}})\slash T) 
\end{align}
The number densities of particles and antiparticles can be found in the normal manner, using
\begin{align}
n_p &= \int \dfrac{d^3p}{(2\pi)^3} \exp(-(E-E_0-\mu_p)\slash T) \nonumber \\
n_{\bar{p}} &= \int \dfrac{d^3p}{(2\pi)^3} \exp(-(E+E_0-\mu_{\bar{p}})\slash T) \end{align}
If we use the non-relativistic relation $E = p^2 \slash 2 m$, then we find
\begin{align}
\mu_p &= - E_0 + T \ln(\lambda^3n_p) \nonumber \\
\mu_{\bar{p}} &= E_0 + T \ln(\lambda^3 n_{\bar{p}}),
\end{align}
where $\lambda = \sqrt{ 2\pi m T}$.  In the above relation, the first term can be interpretted as an external chemical potential (due to the ``driving" effect of the $\mathcal{O}_6$ operator), while the $T \ln(\lambda^3 n_p)$ is the usual chemical potential of an ideal gas.

If a lepton-number-violating process or baryon-number-violating establishes chemical equilibrium between the species, then the chemical potentials will be equal, $\mu_p = \mu_{\bar{p}}$.  This gives the expected result
\begin{equation}
\dfrac{n_p}{n_{\bar{p}}} = e^{2 E_0 \slash T}.
\end{equation}
A similar result can be derived using the relativistic relation $E = p$ instead.

\section{Calculation of Lepton-Number-Violating Cross Section and Reaction Rate}
\label{sec:cross_section_reaction_rate_apndx}

In this section, we calculate the cross section and reaction rate for the processes shown in Fig.~\ref{fig:lepton_violation}, assuming a thermal number density for Higgs bosons and neutrinos, as discussed in Section~\ref{sec:lepton_number_violating}.  This improves the order of magnitude estimates used in~\cite{Kusenko:2014lra}.  As explained in the text, we can use the approximate cross section with the energy shift due to the $\mathcal{O}_6$ operator set equal to zero.  In this approximation, the reaction rates for processes with neutrinos and antineutrinos are equal.

The top two diagrams of Fig.~\ref{fig:lepton_violation} are the $s$- and $t$-channel diagrams of the process $h^0 \nu \rightarrow h^0 \bar{\nu}$, whereas the bottom diagram describes the process $\nu \nu \rightarrow h^0 h^0$.  The $s$-channel has a resonance at $E \sim M_R$; however, the typical energy scale $T$ is far beneath this.  For completeness, we include the resonance, although it will have negligible effect.  In calculating these cross sections, we follow the conventions of \cite{Dreiner:2008tw} for the Feynman rules of Majorana fermions.

The matrix element for the $\nu_\ell h^0 \rightarrow \bar{\nu}_{L} h^0$ process is
\begin{widetext}
\begin{align}
- i \mathcal{M} &= i\sum_i  \dfrac{Y_{Li} Y_{i\ell}^*}{2} \left[ \dfrac{M_{Ri} - i \Gamma_i \slash 2}{s - M_{Ri}^2 + i \Gamma_i M_{Ri} + \Gamma_i^2 \slash 4}  + \dfrac{M_{Ri}- i \Gamma_i \slash 2}{t - M_{Ri}^2 + i \Gamma_i M_{Ri} + \Gamma_i^2 \slash 4}  \right] x_{L\alpha}(p_1,s_1) y^{\beta}_\ell(p_4,s_4) \delta_\beta^\alpha,
\end{align}
\end{widetext}
where $s$ and $t$ are the Mandelstam variables, and $\Gamma_i$ is the width of the right-handed Majorana neutrino.  (For a discussion of Breit-Wignar propagators, see \cite{Nowakowski:1993iu}).  The indices 1, 2, 3, and 4 refer to the incoming neutrino, incoming Higgs boson, outgoing Higgs boson, and outgoing antineutrino, in that order.  The index $i$ indicates a sum over the heavy right-handed Majorana neutrinos.  Let us define
\begin{align}
A_i &= s - M_{Ri}^2 + \Gamma_i^2 \slash 4, \nonumber \\
B_i &= t - M_{Ri}^2 + \Gamma_i^2 \slash 4, \nonumber \\
C_i &= \Gamma_i M_{Ri}.
\end{align}

Then the matrix element squared, summed over both the initial and final spin states (as discussed in \cite{Giudice:2003jh}), is
\begin{align}
&\sum_{s_1,s_2} |\mathcal{M}|^2  =\sum_i 2 p_1 \times p_4 \dfrac{|Y_{Li}|^2 |Y_{i\ell}|^2}{4} \left(M_{Ri}^2 + \dfrac{\Gamma_i^2 }{ 4} \right) \times \nonumber \\
&  \left[ \dfrac{1}{A_i^2 + C_i^2} + \dfrac{1}{B_i^2 + C_i^2} + \dfrac{2 (A_i B_i + C_i^2)}{(A_i B_i + C_i^2)^2 + C_i^2(A_i - B_i)^2} \right].
\end{align}
In the center of mass reference frame, the cross section is 
\begin{widetext}
\begin{align}
 \sigma_{CM}(s) 
= \dfrac{1}{16 \pi}  \sum_i \dfrac{|Y_{Li}|^2 |Y_{i\ell}|^2}{4} \left( M_{Ri}^2 + \dfrac{\Gamma_i}{4} \right) \int_{-s}^0 dt \, (s+t) \left[ \dfrac{1}{A_i^2 + C_i^2} + \dfrac{1}{B_i^2 + C_i^2} + \dfrac{2 (A_i B_i + C_i^2)}{(A_i B_i + C_i^2)^2 + C_i^2(A_i - B_i)^2} \right].
\end{align}

Generically, the thermally averaged cross section is related to the CM cross section by~\cite{Cannoni:2013zya}
\begin{align}
&\left< \sigma v \right> = \dfrac{1}{8T \times m_1^2 K_2(m_1 \slash T) \times m_2^2 K_2(m_2 \slash T)} \int_{(m_1 + m_2)^2}^\infty \dfrac{[s - (m_1 - m_2)^2] [s - (m_1 + m_2)^2]}{\sqrt{s}}  K_1(\sqrt{s} \slash T) \sigma_{CM}(s),
\label{eq:cross_section}
\end{align}
and so the thermally averaged cross section for $h^0 \nu \rightarrow h^0 \bar{\nu}$ is
\begin{align}
&\left< \sigma(h^0 \nu \rightarrow h^0 \bar{\nu}) v \right> = \sum_i \dfrac{ |Y_{Li}|^2 |Y_{i\ell}|^2 }{512 \pi} \left( M_{Ri}^2 + \dfrac{\Gamma_i}{4} \right)  \int_0^x dx  
\int_0^x dy  (x^2 - y^2) K_1(x)   \left[ \dfrac{1}{(x^2 T^2 - M_{Ri}^2 + \Gamma_i^2 \slash 4)^2 + \Gamma_i^2 M_{Ri}^2}  \right.\nonumber \\
&  \left. + \dfrac{1}{(y^2 T^2 + M_{Ri}^2 - \Gamma_i^2 \slash 4)^2 + \Gamma_i^2 M_{Ri}^2} - \dfrac{2 ((x^2 T^2 - M_{Ri}^2 + \Gamma_i^2 \slash 4) (y^2 T^2 + M_{Ri}^2 - \Gamma_i^2 \slash 4) -\Gamma_i^2 M_{Ri}^2)}{((x^2 T^2 - M_{Ri}^2 + \Gamma_i^2 \slash 4)(y^2 T^2 + M_{Ri}^2 - \Gamma_i^2 \slash 4) - \Gamma_i^2 M_{Ri}^2)^2 + \Gamma_i^2 M_{Ri}^2 (x^2 +y^2)^2 T^4} \right]
\end{align}
\end{widetext}
where we have introduced the dimensionless variables $x \equiv \sqrt{s} \slash T$ and $y \equiv \sqrt{-t} \slash T$.  Since the temperature evolves in time, the cross section also does; however, when expanded in powers of $T \slash M_{Ri}$, the lowest order contribution is $\sim 1 \slash M_{Ri}^2$, as expected.  Repeating the same steps with the $\nu_\ell \nu_L \rightarrow h^0 h^0$ cross section, which does not have a resonance, gives 
\begin{align}
\left< \sigma(\nu_\ell \nu_L \rightarrow h^0 h^0) v \right> 
&= \sum_i \dfrac{|Y_{Li}|^2 |Y_{i\ell}|^2}{64 \pi M_{Ri}^2}.
\end{align}

The reaction rates are related to these cross sections by
\begin{equation}
\gamma^{eq}(\alpha \beta \rightarrow \gamma \delta) = (n_\alpha^{eq}) (n_\beta^{eq}) \left< \sigma(\alpha \beta \rightarrow \gamma \delta) v \right>,
\label{eq:gamma_to_sigma}
\end{equation}
which holds for any $2 \rightarrow 2$ process.  Since we take $E_0 = 0$ in this section, the number densities for Higgs bosons, neutrinos, and antineutrinos are all equal to $T^3 \slash \pi^2$, and so for both processes, 
\begin{align}
\gamma^0 & = \dfrac{T^6}{\pi^4} \left< \sigma v \right>.
\end{align}

As noted in the text, in order to simplify the calculation, we will consider only the case in which the flavor indices $\ell$ and $L$ are equal, and the contribution of a single right-handed neutrino dominates.  Its decay rate is $\Gamma \sim y^2 M_R \slash 16 \pi$, from the only decay $N_R \rightarrow h^0 \nu_L$.

\bibliographystyle{apsrev4-1}
\bibliography{Reference}

\begin{thebibliography}{68}%
\makeatletter
\providecommand \@ifxundefined [1]{%
 \@ifx{#1\undefined}
}%
\providecommand \@ifnum [1]{%
 \ifnum #1\expandafter \@firstoftwo
 \else \expandafter \@secondoftwo
 \fi
}%
\providecommand \@ifx [1]{%
 \ifx #1\expandafter \@firstoftwo
 \else \expandafter \@secondoftwo
 \fi
}%
\providecommand \natexlab [1]{#1}%
\providecommand \enquote  [1]{``#1''}%
\providecommand \bibnamefont  [1]{#1}%
\providecommand \bibfnamefont [1]{#1}%
\providecommand \citenamefont [1]{#1}%
\providecommand \href@noop [0]{\@secondoftwo}%
\providecommand \href [0]{\begingroup \@sanitize@url \@href}%
\providecommand \@href[1]{\@@startlink{#1}\@@href}%
\providecommand \@@href[1]{\endgroup#1\@@endlink}%
\providecommand \@sanitize@url [0]{\catcode `\\12\catcode `\$12\catcode
  `\&12\catcode `\#12\catcode `\^12\catcode `\_12\catcode `\%12\relax}%
\providecommand \@@startlink[1]{}%
\providecommand \@@endlink[0]{}%
\providecommand \url  [0]{\begingroup\@sanitize@url \@url }%
\providecommand \@url [1]{\endgroup\@href {#1}{\urlprefix }}%
\providecommand \urlprefix  [0]{URL }%
\providecommand \Eprint [0]{\href }%
\providecommand \doibase [0]{http://dx.doi.org/}%
\providecommand \selectlanguage [0]{\@gobble}%
\providecommand \bibinfo  [0]{\@secondoftwo}%
\providecommand \bibfield  [0]{\@secondoftwo}%
\providecommand \translation [1]{[#1]}%
\providecommand \BibitemOpen [0]{}%
\providecommand \bibitemStop [0]{}%
\providecommand \bibitemNoStop [0]{.\EOS\space}%
\providecommand \EOS [0]{\spacefactor3000\relax}%
\providecommand \BibitemShut  [1]{\csname bibitem#1\endcsname}%
\let\auto@bib@innerbib\@empty
\bibitem [{\citenamefont {Kusenko}\ \emph {et~al.}(2015)\citenamefont
  {Kusenko}, \citenamefont {Pearce},\ and\ \citenamefont
  {Yang}}]{Kusenko:2014lra}%
  \BibitemOpen
  \bibfield  {author} {\bibinfo {author} {\bibfnamefont {A.}~\bibnamefont
  {Kusenko}}, \bibinfo {author} {\bibfnamefont {L.}~\bibnamefont {Pearce}}, \
  and\ \bibinfo {author} {\bibfnamefont {L.}~\bibnamefont {Yang}},\ }\href
  {\doibase 10.1103/PhysRevLett.114.061302} {\bibfield  {journal} {\bibinfo
  {journal} {Phys. Rev. Lett.}\ }\textbf {\bibinfo {volume} {114}},\ \bibinfo
  {pages} {061302} (\bibinfo {year} {2015})},\ \Eprint
  {http://arxiv.org/abs/1410.0722} {arXiv:1410.0722 [hep-ph]} \BibitemShut
  {NoStop}%
\bibitem [{\citenamefont {Bunch}\ and\ \citenamefont
  {Davies}(1978)}]{Bunch:1978yq}%
  \BibitemOpen
  \bibfield  {author} {\bibinfo {author} {\bibfnamefont {T.}~\bibnamefont
  {Bunch}}\ and\ \bibinfo {author} {\bibfnamefont {P.}~\bibnamefont {Davies}},\
  }\href {\doibase 10.1098/rspa.1978.0060} {\bibfield  {journal} {\bibinfo
  {journal} {Proc. Roy. Soc. Lond.}\ }\textbf {\bibinfo {volume} {A360}},\
  \bibinfo {pages} {117} (\bibinfo {year} {1978})}\BibitemShut {NoStop}%
\bibitem [{\citenamefont {Hawking}\ and\ \citenamefont
  {Moss}(1982)}]{Hawking:1981fz}%
  \BibitemOpen
  \bibfield  {author} {\bibinfo {author} {\bibfnamefont {S.}~\bibnamefont
  {Hawking}}\ and\ \bibinfo {author} {\bibfnamefont {I.}~\bibnamefont {Moss}},\
  }\href {\doibase 10.1016/0370-2693(82)90946-7} {\bibfield  {journal}
  {\bibinfo  {journal} {Phys. Lett.}\ }\textbf {\bibinfo {volume} {B110}},\
  \bibinfo {pages} {35} (\bibinfo {year} {1982})}\BibitemShut {NoStop}%
\bibitem [{\citenamefont {Linde}(1982)}]{Linde:1982uu}%
  \BibitemOpen
  \bibfield  {author} {\bibinfo {author} {\bibfnamefont {A.~D.}\ \bibnamefont
  {Linde}},\ }\href {\doibase 10.1016/0370-2693(82)90293-3} {\bibfield
  {journal} {\bibinfo  {journal} {Phys.Lett.}\ }\textbf {\bibinfo {volume}
  {B116}},\ \bibinfo {pages} {335} (\bibinfo {year} {1982})}\BibitemShut
  {NoStop}%
\bibitem [{\citenamefont {Starobinsky}(1982)}]{Starobinsky:1982ee}%
  \BibitemOpen
  \bibfield  {author} {\bibinfo {author} {\bibfnamefont {A.~A.}\ \bibnamefont
  {Starobinsky}},\ }\href {\doibase 10.1016/0370-2693(82)90541-X} {\bibfield
  {journal} {\bibinfo  {journal} {Phys.Lett.}\ }\textbf {\bibinfo {volume}
  {B117}},\ \bibinfo {pages} {175} (\bibinfo {year} {1982})}\BibitemShut
  {NoStop}%
\bibitem [{\citenamefont {Vilenkin}\ and\ \citenamefont
  {Ford}(1982)}]{Vilenkin:1982wt}%
  \BibitemOpen
  \bibfield  {author} {\bibinfo {author} {\bibfnamefont {A.}~\bibnamefont
  {Vilenkin}}\ and\ \bibinfo {author} {\bibfnamefont {L.~H.}\ \bibnamefont
  {Ford}},\ }\href {\doibase 10.1103/PhysRevD.26.1231} {\bibfield  {journal}
  {\bibinfo  {journal} {Phys.Rev.}\ }\textbf {\bibinfo {volume} {D26}},\
  \bibinfo {pages} {1231} (\bibinfo {year} {1982})}\BibitemShut {NoStop}%
\bibitem [{\citenamefont {Starobinsky}\ and\ \citenamefont
  {Yokoyama}(1994)}]{Starobinsky:1994bd}%
  \BibitemOpen
  \bibfield  {author} {\bibinfo {author} {\bibfnamefont {A.~A.}\ \bibnamefont
  {Starobinsky}}\ and\ \bibinfo {author} {\bibfnamefont {J.}~\bibnamefont
  {Yokoyama}},\ }\href {\doibase 10.1103/PhysRevD.50.6357} {\bibfield
  {journal} {\bibinfo  {journal} {Phys. Rev.}\ }\textbf {\bibinfo {volume}
  {D50}},\ \bibinfo {pages} {6357} (\bibinfo {year} {1994})},\ \Eprint
  {http://arxiv.org/abs/astro-ph/9407016} {arXiv:astro-ph/9407016 [astro-ph]}
  \BibitemShut {NoStop}%
\bibitem [{\citenamefont {Enqvist}\ \emph {et~al.}(2013)\citenamefont
  {Enqvist}, \citenamefont {Meriniemi},\ and\ \citenamefont
  {Nurmi}}]{Enqvist:2013kaa}%
  \BibitemOpen
  \bibfield  {author} {\bibinfo {author} {\bibfnamefont {K.}~\bibnamefont
  {Enqvist}}, \bibinfo {author} {\bibfnamefont {T.}~\bibnamefont {Meriniemi}},
  \ and\ \bibinfo {author} {\bibfnamefont {S.}~\bibnamefont {Nurmi}},\ }\href
  {\doibase 10.1088/1475-7516/2013/10/057} {\bibfield  {journal} {\bibinfo
  {journal} {JCAP}\ }\textbf {\bibinfo {volume} {1310}},\ \bibinfo {pages}
  {057} (\bibinfo {year} {2013})},\ \Eprint {http://arxiv.org/abs/1306.4511}
  {arXiv:1306.4511 [hep-ph]} \BibitemShut {NoStop}%
\bibitem [{\citenamefont {Enqvist}\ \emph
  {et~al.}(2014{\natexlab{a}})\citenamefont {Enqvist}, \citenamefont
  {Meriniemi},\ and\ \citenamefont {Nurmi}}]{Enqvist:2014bua}%
  \BibitemOpen
  \bibfield  {author} {\bibinfo {author} {\bibfnamefont {K.}~\bibnamefont
  {Enqvist}}, \bibinfo {author} {\bibfnamefont {T.}~\bibnamefont {Meriniemi}},
  \ and\ \bibinfo {author} {\bibfnamefont {S.}~\bibnamefont {Nurmi}},\ }\href
  {\doibase 10.1088/1475-7516/2014/07/025} {\bibfield  {journal} {\bibinfo
  {journal} {JCAP}\ }\textbf {\bibinfo {volume} {1407}},\ \bibinfo {pages}
  {025} (\bibinfo {year} {2014}{\natexlab{a}})},\ \Eprint
  {http://arxiv.org/abs/1404.3699} {arXiv:1404.3699 [hep-ph]} \BibitemShut
  {NoStop}%
\bibitem [{\citenamefont {Sakharov}(1967)}]{Sakharov:1967dj}%
  \BibitemOpen
  \bibfield  {author} {\bibinfo {author} {\bibfnamefont {A.}~\bibnamefont
  {Sakharov}},\ }\href {\doibase 10.1070/PU1991v034n05ABEH002497} {\bibfield
  {journal} {\bibinfo  {journal} {Pisma Zh. Eksp. Teor. Fiz.}\ }\textbf
  {\bibinfo {volume} {5}},\ \bibinfo {pages} {32} (\bibinfo {year}
  {1967})}\BibitemShut {NoStop}%
\bibitem [{\citenamefont {Dine}\ \emph {et~al.}(1991)\citenamefont {Dine},
  \citenamefont {Huet}, \citenamefont {Singleton},\ and\ \citenamefont
  {Susskind}}]{Dine:1990fj}%
  \BibitemOpen
  \bibfield  {author} {\bibinfo {author} {\bibfnamefont {M.}~\bibnamefont
  {Dine}}, \bibinfo {author} {\bibfnamefont {P.}~\bibnamefont {Huet}}, \bibinfo
  {author} {\bibfnamefont {R.~L.}\ \bibnamefont {Singleton}}, \ and\ \bibinfo
  {author} {\bibfnamefont {L.}~\bibnamefont {Susskind}},\ }\href {\doibase
  10.1016/0370-2693(91)91905-B} {\bibfield  {journal} {\bibinfo  {journal}
  {Phys. Lett.}\ }\textbf {\bibinfo {volume} {B257}},\ \bibinfo {pages} {351}
  (\bibinfo {year} {1991})}\BibitemShut {NoStop}%
\bibitem [{\citenamefont {Pearce}\ \emph {et~al.}(2015)\citenamefont {Pearce},
  \citenamefont {Yang}, \citenamefont {Kusenko},\ and\ \citenamefont
  {Peloso}}]{Pearce:2015nga}%
  \BibitemOpen
  \bibfield  {author} {\bibinfo {author} {\bibfnamefont {L.}~\bibnamefont
  {Pearce}}, \bibinfo {author} {\bibfnamefont {L.}~\bibnamefont {Yang}},
  \bibinfo {author} {\bibfnamefont {A.}~\bibnamefont {Kusenko}}, \ and\
  \bibinfo {author} {\bibfnamefont {M.}~\bibnamefont {Peloso}},\ }\href
  {\doibase 10.1103/PhysRevD.92.023509} {\bibfield  {journal} {\bibinfo
  {journal} {Phys. Rev.}\ }\textbf {\bibinfo {volume} {D92}},\ \bibinfo {pages}
  {023509} (\bibinfo {year} {2015})},\ \Eprint
  {http://arxiv.org/abs/1505.02461} {arXiv:1505.02461 [hep-ph]} \BibitemShut
  {NoStop}%
\bibitem [{\citenamefont {Chiba}\ \emph {et~al.}(2004)\citenamefont {Chiba},
  \citenamefont {Takahashi},\ and\ \citenamefont {Yamaguchi}}]{Chiba:2003vp}%
  \BibitemOpen
  \bibfield  {author} {\bibinfo {author} {\bibfnamefont {T.}~\bibnamefont
  {Chiba}}, \bibinfo {author} {\bibfnamefont {F.}~\bibnamefont {Takahashi}}, \
  and\ \bibinfo {author} {\bibfnamefont {M.}~\bibnamefont {Yamaguchi}},\ }\href
  {\doibase 10.1103/PhysRevLett.92.011301} {\bibfield  {journal} {\bibinfo
  {journal} {Phys. Rev. Lett.}\ }\textbf {\bibinfo {volume} {92}},\ \bibinfo
  {pages} {011301} (\bibinfo {year} {2004})},\ \Eprint
  {http://arxiv.org/abs/hep-ph/0304102} {arXiv:hep-ph/0304102 [hep-ph]}
  \BibitemShut {NoStop}%
\bibitem [{\citenamefont {Kusenko}\ \emph {et~al.}(2014)\citenamefont
  {Kusenko}, \citenamefont {Schmitz},\ and\ \citenamefont
  {Yanagida}}]{Kusenko:2014uta}%
  \BibitemOpen
  \bibfield  {author} {\bibinfo {author} {\bibfnamefont {A.}~\bibnamefont
  {Kusenko}}, \bibinfo {author} {\bibfnamefont {K.}~\bibnamefont {Schmitz}}, \
  and\ \bibinfo {author} {\bibfnamefont {T.~T.}\ \bibnamefont {Yanagida}},\
  }\href@noop {} {\  (\bibinfo {year} {2014})},\ \Eprint
  {http://arxiv.org/abs/1412.2043} {arXiv:1412.2043 [hep-ph]} \BibitemShut
  {NoStop}%
\bibitem [{\citenamefont {{Peebles}}(1987{\natexlab{a}})}]{Peebles1987}%
  \BibitemOpen
  \bibfield  {author} {\bibinfo {author} {\bibfnamefont {P.}~\bibnamefont
  {{Peebles}}},\ }\href {\doibase 10.1038/327210a0} {\bibfield  {journal}
  {\bibinfo  {journal} {Nature}\ }\textbf {\bibinfo {volume} {327}},\ \bibinfo
  {pages} {210} (\bibinfo {year} {1987}{\natexlab{a}})}\BibitemShut {NoStop}%
\bibitem [{\citenamefont
  {{Peebles}}(1987{\natexlab{b}})}]{1987ApJ...315L..73P}%
  \BibitemOpen
  \bibfield  {author} {\bibinfo {author} {\bibfnamefont {P.}~\bibnamefont
  {{Peebles}}},\ }\href {\doibase 10.1086/184863} {\bibfield  {journal}
  {\bibinfo  {journal} {Astrophys. J. Lett.}\ }\textbf {\bibinfo {volume}
  {315}},\ \bibinfo {pages} {L73} (\bibinfo {year}
  {1987}{\natexlab{b}})}\BibitemShut {NoStop}%
\bibitem [{\citenamefont {Enqvist}\ and\ \citenamefont
  {McDonald}(1999)}]{Enqvist:1998pf}%
  \BibitemOpen
  \bibfield  {author} {\bibinfo {author} {\bibfnamefont {K.}~\bibnamefont
  {Enqvist}}\ and\ \bibinfo {author} {\bibfnamefont {J.}~\bibnamefont
  {McDonald}},\ }\href {\doibase 10.1103/PhysRevLett.83.2510} {\bibfield
  {journal} {\bibinfo  {journal} {Phys. Rev. Lett.}\ }\textbf {\bibinfo
  {volume} {83}},\ \bibinfo {pages} {2510} (\bibinfo {year} {1999})},\ \Eprint
  {http://arxiv.org/abs/hep-ph/9811412} {arXiv:hep-ph/9811412 [hep-ph]}
  \BibitemShut {NoStop}%
\bibitem [{\citenamefont {Enqvist}\ and\ \citenamefont
  {McDonald}(2000)}]{Enqvist:1999hv}%
  \BibitemOpen
  \bibfield  {author} {\bibinfo {author} {\bibfnamefont {K.}~\bibnamefont
  {Enqvist}}\ and\ \bibinfo {author} {\bibfnamefont {J.}~\bibnamefont
  {McDonald}},\ }\href {\doibase 10.1103/PhysRevD.62.043502} {\bibfield
  {journal} {\bibinfo  {journal} {Phys. Rev.}\ }\textbf {\bibinfo {volume}
  {D62}},\ \bibinfo {pages} {043502} (\bibinfo {year} {2000})},\ \Eprint
  {http://arxiv.org/abs/hep-ph/9912478} {arXiv:hep-ph/9912478 [hep-ph]}
  \BibitemShut {NoStop}%
\bibitem [{\citenamefont {Harigaya}\ \emph {et~al.}(2014)\citenamefont
  {Harigaya}, \citenamefont {Kamada}, \citenamefont {Kawasaki}, \citenamefont
  {Mukaida},\ and\ \citenamefont {Yamada}}]{Harigaya:2014tla}%
  \BibitemOpen
  \bibfield  {author} {\bibinfo {author} {\bibfnamefont {K.}~\bibnamefont
  {Harigaya}}, \bibinfo {author} {\bibfnamefont {A.}~\bibnamefont {Kamada}},
  \bibinfo {author} {\bibfnamefont {M.}~\bibnamefont {Kawasaki}}, \bibinfo
  {author} {\bibfnamefont {K.}~\bibnamefont {Mukaida}}, \ and\ \bibinfo
  {author} {\bibfnamefont {M.}~\bibnamefont {Yamada}},\ }\href {\doibase
  10.1103/PhysRevD.90.043510} {\bibfield  {journal} {\bibinfo  {journal} {Phys.
  Rev.}\ }\textbf {\bibinfo {volume} {D90}},\ \bibinfo {pages} {043510}
  (\bibinfo {year} {2014})},\ \Eprint {http://arxiv.org/abs/1404.3138}
  {arXiv:1404.3138 [hep-ph]} \BibitemShut {NoStop}%
\bibitem [{\citenamefont {Ade}\ \emph {et~al.}(2014)\citenamefont {Ade} \emph
  {et~al.}}]{Ade:2013uln}%
  \BibitemOpen
  \bibfield  {author} {\bibinfo {author} {\bibfnamefont {P.}~\bibnamefont
  {Ade}} \emph {et~al.} (\bibinfo {collaboration} {Planck Collaboration}),\
  }\href {\doibase 10.1051/0004-6361/201321569} {\bibfield  {journal} {\bibinfo
   {journal} {Astron. Astrophys.}\ }\textbf {\bibinfo {volume} {571}},\
  \bibinfo {pages} {A22} (\bibinfo {year} {2014})},\ \Eprint
  {http://arxiv.org/abs/1303.5082} {arXiv:1303.5082 [astro-ph.CO]} \BibitemShut
  {NoStop}%
\bibitem [{\citenamefont {Degrassi}\ \emph {et~al.}(2012)\citenamefont
  {Degrassi}, \citenamefont {Di~Vita}, \citenamefont {Elias-Miro},
  \citenamefont {Espinosa}, \citenamefont {Giudice}, \citenamefont {Isidori},\
  and\ \citenamefont {Strumia}}]{Degrassi:2012ry}%
  \BibitemOpen
  \bibfield  {author} {\bibinfo {author} {\bibfnamefont {G.}~\bibnamefont
  {Degrassi}}, \bibinfo {author} {\bibfnamefont {S.}~\bibnamefont {Di~Vita}},
  \bibinfo {author} {\bibfnamefont {J.}~\bibnamefont {Elias-Miro}}, \bibinfo
  {author} {\bibfnamefont {J.~R.}\ \bibnamefont {Espinosa}}, \bibinfo {author}
  {\bibfnamefont {G.~F.}\ \bibnamefont {Giudice}}, \bibinfo {author}
  {\bibfnamefont {G.}~\bibnamefont {Isidori}}, \ and\ \bibinfo {author}
  {\bibfnamefont {A.}~\bibnamefont {Strumia}},\ }\href {\doibase
  10.1007/JHEP08(2012)098} {\bibfield  {journal} {\bibinfo  {journal} {JHEP}\
  }\textbf {\bibinfo {volume} {1208}},\ \bibinfo {pages} {098} (\bibinfo {year}
  {2012})},\ \Eprint {http://arxiv.org/abs/1205.6497} {arXiv:1205.6497
  [hep-ph]} \BibitemShut {NoStop}%
\bibitem [{\citenamefont {Kusenko}\ and\ \citenamefont
  {Langacker}(1997)}]{Kusenko:1996xt}%
  \BibitemOpen
  \bibfield  {author} {\bibinfo {author} {\bibfnamefont {A.}~\bibnamefont
  {Kusenko}}\ and\ \bibinfo {author} {\bibfnamefont {P.}~\bibnamefont
  {Langacker}},\ }\href {\doibase 10.1016/S0370-2693(96)01470-0} {\bibfield
  {journal} {\bibinfo  {journal} {Phys. Lett.}\ }\textbf {\bibinfo {volume}
  {B391}},\ \bibinfo {pages} {29} (\bibinfo {year} {1997})},\ \Eprint
  {http://arxiv.org/abs/hep-ph/9608340} {arXiv:hep-ph/9608340 [hep-ph]}
  \BibitemShut {NoStop}%
\bibitem [{\citenamefont {Kusenko}\ \emph {et~al.}(1996)\citenamefont
  {Kusenko}, \citenamefont {Langacker},\ and\ \citenamefont
  {Segre}}]{Kusenko:1996jn}%
  \BibitemOpen
  \bibfield  {author} {\bibinfo {author} {\bibfnamefont {A.}~\bibnamefont
  {Kusenko}}, \bibinfo {author} {\bibfnamefont {P.}~\bibnamefont {Langacker}},
  \ and\ \bibinfo {author} {\bibfnamefont {G.}~\bibnamefont {Segre}},\ }\href
  {\doibase 10.1103/PhysRevD.54.5824} {\bibfield  {journal} {\bibinfo
  {journal} {Phys. Rev.}\ }\textbf {\bibinfo {volume} {D54}},\ \bibinfo {pages}
  {5824} (\bibinfo {year} {1996})},\ \Eprint
  {http://arxiv.org/abs/hep-ph/9602414} {arXiv:hep-ph/9602414 [hep-ph]}
  \BibitemShut {NoStop}%
\bibitem [{\citenamefont {Kobakhidze}\ and\ \citenamefont
  {Spencer-Smith}(2013)}]{Kobakhidze:2013tn}%
  \BibitemOpen
  \bibfield  {author} {\bibinfo {author} {\bibfnamefont {A.}~\bibnamefont
  {Kobakhidze}}\ and\ \bibinfo {author} {\bibfnamefont {A.}~\bibnamefont
  {Spencer-Smith}},\ }\href {\doibase 10.1016/j.physletb.2013.04.013}
  {\bibfield  {journal} {\bibinfo  {journal} {Phys. Lett.}\ }\textbf {\bibinfo
  {volume} {B722}},\ \bibinfo {pages} {130} (\bibinfo {year} {2013})},\ \Eprint
  {http://arxiv.org/abs/1301.2846} {arXiv:1301.2846 [hep-ph]} \BibitemShut
  {NoStop}%
\bibitem [{\citenamefont {Dai}\ \emph {et~al.}(2014)\citenamefont {Dai},
  \citenamefont {Kamionkowski},\ and\ \citenamefont {Wang}}]{Dai:2014jja}%
  \BibitemOpen
  \bibfield  {author} {\bibinfo {author} {\bibfnamefont {L.}~\bibnamefont
  {Dai}}, \bibinfo {author} {\bibfnamefont {M.}~\bibnamefont {Kamionkowski}}, \
  and\ \bibinfo {author} {\bibfnamefont {J.}~\bibnamefont {Wang}},\ }\href
  {\doibase 10.1103/PhysRevLett.113.041302} {\bibfield  {journal} {\bibinfo
  {journal} {Phys. Rev. Lett.}\ }\textbf {\bibinfo {volume} {113}},\ \bibinfo
  {pages} {041302} (\bibinfo {year} {2014})},\ \Eprint
  {http://arxiv.org/abs/1404.6704} {arXiv:1404.6704 [astro-ph.CO]} \BibitemShut
  {NoStop}%
\bibitem [{\citenamefont {Kolb}\ and\ \citenamefont
  {Turner}(1990)}]{Kolb:1990vq}%
  \BibitemOpen
  \bibfield  {author} {\bibinfo {author} {\bibfnamefont {E.~W.}\ \bibnamefont
  {Kolb}}\ and\ \bibinfo {author} {\bibfnamefont {M.~S.}\ \bibnamefont
  {Turner}},\ }\href@noop {} {\emph {\bibinfo {title} {The Early Universe,
  Frontiers in Physics lecture note series}}},\ Vol.~\bibinfo {volume} {69}\
  (\bibinfo  {publisher} {Addison-Wesley},\ \bibinfo {address} {Reading, MA},\
  \bibinfo {year} {1990})\BibitemShut {NoStop}%
\bibitem [{\citenamefont {Andreassen}\ \emph {et~al.}(2014)\citenamefont
  {Andreassen}, \citenamefont {Frost},\ and\ \citenamefont
  {Schwartz}}]{Andreassen:2014gha}%
  \BibitemOpen
  \bibfield  {author} {\bibinfo {author} {\bibfnamefont {A.}~\bibnamefont
  {Andreassen}}, \bibinfo {author} {\bibfnamefont {W.}~\bibnamefont {Frost}}, \
  and\ \bibinfo {author} {\bibfnamefont {M.~D.}\ \bibnamefont {Schwartz}},\
  }\href {\doibase 10.1103/PhysRevLett.113.241801} {\bibfield  {journal}
  {\bibinfo  {journal} {Phys.Rev.Lett.}\ }\textbf {\bibinfo {volume} {113}},\
  \bibinfo {pages} {241801} (\bibinfo {year} {2014})},\ \Eprint
  {http://arxiv.org/abs/1408.0292} {arXiv:1408.0292 [hep-ph]} \BibitemShut
  {NoStop}%
\bibitem [{\citenamefont {Andreassen}\ \emph {et~al.}(2015)\citenamefont
  {Andreassen}, \citenamefont {Frost},\ and\ \citenamefont
  {Schwartz}}]{Andreassen:2014eha}%
  \BibitemOpen
  \bibfield  {author} {\bibinfo {author} {\bibfnamefont {A.}~\bibnamefont
  {Andreassen}}, \bibinfo {author} {\bibfnamefont {W.}~\bibnamefont {Frost}}, \
  and\ \bibinfo {author} {\bibfnamefont {M.~D.}\ \bibnamefont {Schwartz}},\
  }\href {\doibase 10.1103/PhysRevD.91.016009} {\bibfield  {journal} {\bibinfo
  {journal} {Phys.Rev.}\ }\textbf {\bibinfo {volume} {D91}},\ \bibinfo {pages}
  {016009} (\bibinfo {year} {2015})},\ \Eprint {http://arxiv.org/abs/1408.0287}
  {arXiv:1408.0287 [hep-ph]} \BibitemShut {NoStop}%
\bibitem [{\citenamefont {Nielsen}(1975)}]{Nielsen:1975fs}%
  \BibitemOpen
  \bibfield  {author} {\bibinfo {author} {\bibfnamefont {N.}~\bibnamefont
  {Nielsen}},\ }\href {\doibase 10.1016/0550-3213(75)90301-6} {\bibfield
  {journal} {\bibinfo  {journal} {Nucl.Phys.}\ }\textbf {\bibinfo {volume}
  {B101}},\ \bibinfo {pages} {173} (\bibinfo {year} {1975})}\BibitemShut
  {NoStop}%
\bibitem [{\citenamefont {Fukuda}\ and\ \citenamefont
  {Kugo}(1976)}]{Fukuda:1975di}%
  \BibitemOpen
  \bibfield  {author} {\bibinfo {author} {\bibfnamefont {R.}~\bibnamefont
  {Fukuda}}\ and\ \bibinfo {author} {\bibfnamefont {T.}~\bibnamefont {Kugo}},\
  }\href {\doibase 10.1103/PhysRevD.13.3469} {\bibfield  {journal} {\bibinfo
  {journal} {Phys.Rev.}\ }\textbf {\bibinfo {volume} {D13}},\ \bibinfo {pages}
  {3469} (\bibinfo {year} {1976})}\BibitemShut {NoStop}%
\bibitem [{\citenamefont {Aitchison}\ and\ \citenamefont
  {Fraser}(1984)}]{Aitchison:1983ns}%
  \BibitemOpen
  \bibfield  {author} {\bibinfo {author} {\bibfnamefont {I.}~\bibnamefont
  {Aitchison}}\ and\ \bibinfo {author} {\bibfnamefont {C.}~\bibnamefont
  {Fraser}},\ }\href {\doibase 10.1016/0003-4916(84)90209-4} {\bibfield
  {journal} {\bibinfo  {journal} {Annals Phys.}\ }\textbf {\bibinfo {volume}
  {156}},\ \bibinfo {pages} {1} (\bibinfo {year} {1984})}\BibitemShut {NoStop}%
\bibitem [{\citenamefont {Di~Luzio}\ and\ \citenamefont
  {Mihaila}(2014)}]{DiLuzio:2014bua}%
  \BibitemOpen
  \bibfield  {author} {\bibinfo {author} {\bibfnamefont {L.}~\bibnamefont
  {Di~Luzio}}\ and\ \bibinfo {author} {\bibfnamefont {L.}~\bibnamefont
  {Mihaila}},\ }\href {\doibase 10.1007/JHEP06(2014)079} {\bibfield  {journal}
  {\bibinfo  {journal} {JHEP}\ }\textbf {\bibinfo {volume} {1406}},\ \bibinfo
  {pages} {079} (\bibinfo {year} {2014})},\ \Eprint
  {http://arxiv.org/abs/1404.7450} {arXiv:1404.7450 [hep-ph]} \BibitemShut
  {NoStop}%
\bibitem [{\citenamefont {Casas}\ \emph {et~al.}(1995)\citenamefont {Casas},
  \citenamefont {Espinosa},\ and\ \citenamefont {Quiros}}]{Casas:1994qy}%
  \BibitemOpen
  \bibfield  {author} {\bibinfo {author} {\bibfnamefont {J.}~\bibnamefont
  {Casas}}, \bibinfo {author} {\bibfnamefont {J.}~\bibnamefont {Espinosa}}, \
  and\ \bibinfo {author} {\bibfnamefont {M.}~\bibnamefont {Quiros}},\ }\href
  {\doibase 10.1016/0370-2693(94)01404-Z} {\bibfield  {journal} {\bibinfo
  {journal} {Phys. Lett.}\ }\textbf {\bibinfo {volume} {B342}},\ \bibinfo
  {pages} {171} (\bibinfo {year} {1995})},\ \Eprint
  {http://arxiv.org/abs/hep-ph/9409458} {arXiv:hep-ph/9409458 [hep-ph]}
  \BibitemShut {NoStop}%
\bibitem [{\citenamefont {Anderson}\ and\ \citenamefont
  {Hall}(1992)}]{Anderson:1991zb}%
  \BibitemOpen
  \bibfield  {author} {\bibinfo {author} {\bibfnamefont {G.~W.}\ \bibnamefont
  {Anderson}}\ and\ \bibinfo {author} {\bibfnamefont {L.~J.}\ \bibnamefont
  {Hall}},\ }\href {\doibase 10.1103/PhysRevD.45.2685} {\bibfield  {journal}
  {\bibinfo  {journal} {Phys. Rev.}\ }\textbf {\bibinfo {volume} {D45}},\
  \bibinfo {pages} {2685} (\bibinfo {year} {1992})}\BibitemShut {NoStop}%
\bibitem [{\citenamefont {Kapusta}\ and\ \citenamefont
  {Gale}(2006)}]{Kapusta:2006pm}%
  \BibitemOpen
  \bibfield  {author} {\bibinfo {author} {\bibfnamefont {J.}~\bibnamefont
  {Kapusta}}\ and\ \bibinfo {author} {\bibfnamefont {C.}~\bibnamefont {Gale}},\
  }\href@noop {} {\emph {\bibinfo {title} {Finite-temperature field theory:
  Principles and applications}}}\ (\bibinfo  {publisher} {Cambridge University
  Press},\ \bibinfo {address} {Cambridge, England},\ \bibinfo {year}
  {2006})\BibitemShut {NoStop}%
\bibitem [{\citenamefont {Enqvist}\ \emph
  {et~al.}(2014{\natexlab{b}})\citenamefont {Enqvist}, \citenamefont {Nurmi},\
  and\ \citenamefont {Rusak}}]{Enqvist:2014tta}%
  \BibitemOpen
  \bibfield  {author} {\bibinfo {author} {\bibfnamefont {K.}~\bibnamefont
  {Enqvist}}, \bibinfo {author} {\bibfnamefont {S.}~\bibnamefont {Nurmi}}, \
  and\ \bibinfo {author} {\bibfnamefont {S.}~\bibnamefont {Rusak}},\ }\href
  {\doibase 10.1088/1475-7516/2014/10/064} {\bibfield  {journal} {\bibinfo
  {journal} {JCAP}\ }\textbf {\bibinfo {volume} {1410}},\ \bibinfo {pages}
  {064} (\bibinfo {year} {2014}{\natexlab{b}})},\ \Eprint
  {http://arxiv.org/abs/1404.3631} {arXiv:1404.3631 [astro-ph.CO]} \BibitemShut
  {NoStop}%
\bibitem [{\citenamefont {Elmfors}\ \emph {et~al.}(1994)\citenamefont
  {Elmfors}, \citenamefont {Enqvist},\ and\ \citenamefont
  {Vilja}}]{Elmfors:1993re}%
  \BibitemOpen
  \bibfield  {author} {\bibinfo {author} {\bibfnamefont {P.}~\bibnamefont
  {Elmfors}}, \bibinfo {author} {\bibfnamefont {K.}~\bibnamefont {Enqvist}}, \
  and\ \bibinfo {author} {\bibfnamefont {I.}~\bibnamefont {Vilja}},\ }\href
  {\doibase 10.1016/0550-3213(94)90512-6} {\bibfield  {journal} {\bibinfo
  {journal} {Nucl.Phys.}\ }\textbf {\bibinfo {volume} {B412}},\ \bibinfo
  {pages} {459} (\bibinfo {year} {1994})},\ \Eprint
  {http://arxiv.org/abs/hep-ph/9307210} {arXiv:hep-ph/9307210 [hep-ph]}
  \BibitemShut {NoStop}%
\bibitem [{\citenamefont {Kofman}\ \emph {et~al.}(1997)\citenamefont {Kofman},
  \citenamefont {Linde},\ and\ \citenamefont {Starobinsky}}]{Kofman:1997yn}%
  \BibitemOpen
  \bibfield  {author} {\bibinfo {author} {\bibfnamefont {L.}~\bibnamefont
  {Kofman}}, \bibinfo {author} {\bibfnamefont {A.~D.}\ \bibnamefont {Linde}}, \
  and\ \bibinfo {author} {\bibfnamefont {A.~A.}\ \bibnamefont {Starobinsky}},\
  }\href {\doibase 10.1103/PhysRevD.56.3258} {\bibfield  {journal} {\bibinfo
  {journal} {Phys.Rev.}\ }\textbf {\bibinfo {volume} {D56}},\ \bibinfo {pages}
  {3258} (\bibinfo {year} {1997})},\ \Eprint
  {http://arxiv.org/abs/hep-ph/9704452} {arXiv:hep-ph/9704452 [hep-ph]}
  \BibitemShut {NoStop}%
\bibitem [{\citenamefont {Garcia-Bellido}\ \emph {et~al.}(2009)\citenamefont
  {Garcia-Bellido}, \citenamefont {Figueroa},\ and\ \citenamefont
  {Rubio}}]{GarciaBellido:2008ab}%
  \BibitemOpen
  \bibfield  {author} {\bibinfo {author} {\bibfnamefont {J.}~\bibnamefont
  {Garcia-Bellido}}, \bibinfo {author} {\bibfnamefont {D.~G.}\ \bibnamefont
  {Figueroa}}, \ and\ \bibinfo {author} {\bibfnamefont {J.}~\bibnamefont
  {Rubio}},\ }\href {\doibase 10.1103/PhysRevD.79.063531} {\bibfield  {journal}
  {\bibinfo  {journal} {Phys.Rev.}\ }\textbf {\bibinfo {volume} {D79}},\
  \bibinfo {pages} {063531} (\bibinfo {year} {2009})},\ \Eprint
  {http://arxiv.org/abs/0812.4624} {arXiv:0812.4624 [hep-ph]} \BibitemShut
  {NoStop}%
\bibitem [{\citenamefont {Figueroa}\ \emph {et~al.}(2015)\citenamefont
  {Figueroa}, \citenamefont {Garcia-Bellido},\ and\ \citenamefont
  {Torrenti}}]{Figueroa:2015rqa}%
  \BibitemOpen
  \bibfield  {author} {\bibinfo {author} {\bibfnamefont {D.~G.}\ \bibnamefont
  {Figueroa}}, \bibinfo {author} {\bibfnamefont {J.}~\bibnamefont
  {Garcia-Bellido}}, \ and\ \bibinfo {author} {\bibfnamefont {F.}~\bibnamefont
  {Torrenti}},\ }\href@noop {} {\  (\bibinfo {year} {2015})},\ \Eprint
  {http://arxiv.org/abs/1504.04600} {arXiv:1504.04600 [astro-ph.CO]}
  \BibitemShut {NoStop}%
\bibitem [{\citenamefont {Garcia-Bellido}\ \emph {et~al.}(2004)\citenamefont
  {Garcia-Bellido}, \citenamefont {Garcia-Perez},\ and\ \citenamefont
  {Gonzalez-Arroyo}}]{GarciaBellido:2003wd}%
  \BibitemOpen
  \bibfield  {author} {\bibinfo {author} {\bibfnamefont {J.}~\bibnamefont
  {Garcia-Bellido}}, \bibinfo {author} {\bibfnamefont {M.}~\bibnamefont
  {Garcia-Perez}}, \ and\ \bibinfo {author} {\bibfnamefont {A.}~\bibnamefont
  {Gonzalez-Arroyo}},\ }\href {\doibase 10.1103/PhysRevD.69.023504} {\bibfield
  {journal} {\bibinfo  {journal} {Phys.Rev.}\ }\textbf {\bibinfo {volume}
  {D69}},\ \bibinfo {pages} {023504} (\bibinfo {year} {2004})},\ \Eprint
  {http://arxiv.org/abs/hep-ph/0304285} {arXiv:hep-ph/0304285 [hep-ph]}
  \BibitemShut {NoStop}%
\bibitem [{\citenamefont {Weldon}(1989)}]{Weldon:1989ys}%
  \BibitemOpen
  \bibfield  {author} {\bibinfo {author} {\bibfnamefont {H.~A.}\ \bibnamefont
  {Weldon}},\ }\href {\doibase 10.1103/PhysRevD.40.2410} {\bibfield  {journal}
  {\bibinfo  {journal} {Phys.Rev.}\ }\textbf {\bibinfo {volume} {D40}},\
  \bibinfo {pages} {2410} (\bibinfo {year} {1989})}\BibitemShut {NoStop}%
\bibitem [{\citenamefont {Enqvist}\ and\ \citenamefont
  {Hogdahl}(2004)}]{Enqvist:2004pr}%
  \BibitemOpen
  \bibfield  {author} {\bibinfo {author} {\bibfnamefont {K.}~\bibnamefont
  {Enqvist}}\ and\ \bibinfo {author} {\bibfnamefont {J.}~\bibnamefont
  {Hogdahl}},\ }\href {\doibase 10.1088/1475-7516/2004/09/013} {\bibfield
  {journal} {\bibinfo  {journal} {JCAP}\ }\textbf {\bibinfo {volume} {0409}},\
  \bibinfo {pages} {013} (\bibinfo {year} {2004})},\ \Eprint
  {http://arxiv.org/abs/hep-ph/0405299} {arXiv:hep-ph/0405299 [hep-ph]}
  \BibitemShut {NoStop}%
\bibitem [{\citenamefont {Drewes}\ and\ \citenamefont
  {Kang}(2013)}]{Drewes:2013iaa}%
  \BibitemOpen
  \bibfield  {author} {\bibinfo {author} {\bibfnamefont {M.}~\bibnamefont
  {Drewes}}\ and\ \bibinfo {author} {\bibfnamefont {J.~U.}\ \bibnamefont
  {Kang}},\ }\href {\doibase 10.1016/j.nuclphysb.2013.07.009,
  10.1016/j.nuclphysb.2014.09.008} {\bibfield  {journal} {\bibinfo  {journal}
  {Nucl.Phys.}\ }\textbf {\bibinfo {volume} {B875}},\ \bibinfo {pages} {315}
  (\bibinfo {year} {2013})},\ \Eprint {http://arxiv.org/abs/1305.0267}
  {arXiv:1305.0267 [hep-ph]} \BibitemShut {NoStop}%
\bibitem [{\citenamefont {Shaposhnikov}(1987)}]{Shaposhnikov:1987tw}%
  \BibitemOpen
  \bibfield  {author} {\bibinfo {author} {\bibfnamefont {M.}~\bibnamefont
  {Shaposhnikov}},\ }\href {\doibase 10.1016/0550-3213(87)90127-1} {\bibfield
  {journal} {\bibinfo  {journal} {Nucl. Phys.}\ }\textbf {\bibinfo {volume}
  {B287}},\ \bibinfo {pages} {757} (\bibinfo {year} {1987})}\BibitemShut
  {NoStop}%
\bibitem [{\citenamefont {Shaposhnikov}(1988)}]{Shaposhnikov:1987pf}%
  \BibitemOpen
  \bibfield  {author} {\bibinfo {author} {\bibfnamefont {M.}~\bibnamefont
  {Shaposhnikov}},\ }\href {\doibase 10.1016/0550-3213(88)90373-2} {\bibfield
  {journal} {\bibinfo  {journal} {Nucl. Phys.}\ }\textbf {\bibinfo {volume}
  {B299}},\ \bibinfo {pages} {797} (\bibinfo {year} {1988})}\BibitemShut
  {NoStop}%
\bibitem [{\citenamefont {Smit}(2004)}]{Smit:2004kh}%
  \BibitemOpen
  \bibfield  {author} {\bibinfo {author} {\bibfnamefont {J.}~\bibnamefont
  {Smit}},\ }\href {\doibase 10.1088/1126-6708/2004/09/067} {\bibfield
  {journal} {\bibinfo  {journal} {JHEP}\ }\textbf {\bibinfo {volume} {0409}},\
  \bibinfo {pages} {067} (\bibinfo {year} {2004})},\ \Eprint
  {http://arxiv.org/abs/hep-ph/0407161} {arXiv:hep-ph/0407161 [hep-ph]}
  \BibitemShut {NoStop}%
\bibitem [{\citenamefont {Brauner}\ \emph {et~al.}(2012)\citenamefont
  {Brauner}, \citenamefont {Taanila}, \citenamefont {Tranberg},\ and\
  \citenamefont {Vuorinen}}]{Brauner:2012gu}%
  \BibitemOpen
  \bibfield  {author} {\bibinfo {author} {\bibfnamefont {T.}~\bibnamefont
  {Brauner}}, \bibinfo {author} {\bibfnamefont {O.}~\bibnamefont {Taanila}},
  \bibinfo {author} {\bibfnamefont {A.}~\bibnamefont {Tranberg}}, \ and\
  \bibinfo {author} {\bibfnamefont {A.}~\bibnamefont {Vuorinen}},\ }\href
  {\doibase 10.1007/JHEP11(2012)076} {\bibfield  {journal} {\bibinfo  {journal}
  {JHEP}\ }\textbf {\bibinfo {volume} {1211}},\ \bibinfo {pages} {076}
  (\bibinfo {year} {2012})},\ \Eprint {http://arxiv.org/abs/1208.5609}
  {arXiv:1208.5609 [hep-ph]} \BibitemShut {NoStop}%
\bibitem [{\citenamefont {Ibe}\ and\ \citenamefont
  {Kaneta}(2015)}]{Ibe:2015nfa}%
  \BibitemOpen
  \bibfield  {author} {\bibinfo {author} {\bibfnamefont {M.}~\bibnamefont
  {Ibe}}\ and\ \bibinfo {author} {\bibfnamefont {K.}~\bibnamefont {Kaneta}},\
  }\href@noop {} {\  (\bibinfo {year} {2015})},\ \Eprint
  {http://arxiv.org/abs/1504.04125} {arXiv:1504.04125 [hep-ph]} \BibitemShut
  {NoStop}%
\bibitem [{\citenamefont {Daido}\ \emph {et~al.}(2015)\citenamefont {Daido},
  \citenamefont {Kitajima},\ and\ \citenamefont {Takahashi}}]{Daido:2015gqa}%
  \BibitemOpen
  \bibfield  {author} {\bibinfo {author} {\bibfnamefont {R.}~\bibnamefont
  {Daido}}, \bibinfo {author} {\bibfnamefont {N.}~\bibnamefont {Kitajima}}, \
  and\ \bibinfo {author} {\bibfnamefont {F.}~\bibnamefont {Takahashi}},\
  }\href@noop {} {\  (\bibinfo {year} {2015})},\ \Eprint
  {http://arxiv.org/abs/1504.07917} {arXiv:1504.07917 [hep-ph]} \BibitemShut
  {NoStop}%
\bibitem [{\citenamefont {Cohen}\ and\ \citenamefont
  {Kaplan}(1987)}]{Cohen:1987vi}%
  \BibitemOpen
  \bibfield  {author} {\bibinfo {author} {\bibfnamefont {A.~G.}\ \bibnamefont
  {Cohen}}\ and\ \bibinfo {author} {\bibfnamefont {D.~B.}\ \bibnamefont
  {Kaplan}},\ }\href {\doibase 10.1016/0370-2693(87)91369-4} {\bibfield
  {journal} {\bibinfo  {journal} {Phys. Lett.}\ }\textbf {\bibinfo {volume}
  {B199}},\ \bibinfo {pages} {251} (\bibinfo {year} {1987})}\BibitemShut
  {NoStop}%
\bibitem [{\citenamefont {Garcia-Bellido}\ \emph {et~al.}(1999)\citenamefont
  {Garcia-Bellido}, \citenamefont {Grigoriev}, \citenamefont {Kusenko},\ and\
  \citenamefont {Shaposhnikov}}]{GarciaBellido:1999sv}%
  \BibitemOpen
  \bibfield  {author} {\bibinfo {author} {\bibfnamefont {J.}~\bibnamefont
  {Garcia-Bellido}}, \bibinfo {author} {\bibfnamefont {D.~{\relax Yu}.}\
  \bibnamefont {Grigoriev}}, \bibinfo {author} {\bibfnamefont {A.}~\bibnamefont
  {Kusenko}}, \ and\ \bibinfo {author} {\bibfnamefont {M.~E.}\ \bibnamefont
  {Shaposhnikov}},\ }\href {\doibase 10.1103/PhysRevD.60.123504} {\bibfield
  {journal} {\bibinfo  {journal} {Phys. Rev.}\ }\textbf {\bibinfo {volume}
  {D60}},\ \bibinfo {pages} {123504} (\bibinfo {year} {1999})},\ \Eprint
  {http://arxiv.org/abs/hep-ph/9902449} {arXiv:hep-ph/9902449 [hep-ph]}
  \BibitemShut {NoStop}%
\bibitem [{\citenamefont {Garcia-Bellido}\ and\ \citenamefont
  {Grigoriev}(2000)}]{GarciaBellido:1999px}%
  \BibitemOpen
  \bibfield  {author} {\bibinfo {author} {\bibfnamefont {J.}~\bibnamefont
  {Garcia-Bellido}}\ and\ \bibinfo {author} {\bibfnamefont {D.~Y.}\
  \bibnamefont {Grigoriev}},\ }\href {\doibase 10.1088/1126-6708/2000/01/017}
  {\bibfield  {journal} {\bibinfo  {journal} {JHEP}\ }\textbf {\bibinfo
  {volume} {0001}},\ \bibinfo {pages} {017} (\bibinfo {year} {2000})},\ \Eprint
  {http://arxiv.org/abs/hep-ph/9912515} {arXiv:hep-ph/9912515 [hep-ph]}
  \BibitemShut {NoStop}%
\bibitem [{\citenamefont {Tranberg}\ and\ \citenamefont
  {Smit}(2003)}]{Tranberg:2003gi}%
  \BibitemOpen
  \bibfield  {author} {\bibinfo {author} {\bibfnamefont {A.}~\bibnamefont
  {Tranberg}}\ and\ \bibinfo {author} {\bibfnamefont {J.}~\bibnamefont
  {Smit}},\ }\href {\doibase 10.1088/1126-6708/2003/11/016} {\bibfield
  {journal} {\bibinfo  {journal} {JHEP}\ }\textbf {\bibinfo {volume} {0311}},\
  \bibinfo {pages} {016} (\bibinfo {year} {2003})},\ \Eprint
  {http://arxiv.org/abs/hep-ph/0310342} {arXiv:hep-ph/0310342 [hep-ph]}
  \BibitemShut {NoStop}%
\bibitem [{\citenamefont {Davoudiasl}\ \emph {et~al.}(2004)\citenamefont
  {Davoudiasl}, \citenamefont {Kitano}, \citenamefont {Kribs}, \citenamefont
  {Murayama},\ and\ \citenamefont {Steinhardt}}]{Davoudiasl:2004gf}%
  \BibitemOpen
  \bibfield  {author} {\bibinfo {author} {\bibfnamefont {H.}~\bibnamefont
  {Davoudiasl}}, \bibinfo {author} {\bibfnamefont {R.}~\bibnamefont {Kitano}},
  \bibinfo {author} {\bibfnamefont {G.~D.}\ \bibnamefont {Kribs}}, \bibinfo
  {author} {\bibfnamefont {H.}~\bibnamefont {Murayama}}, \ and\ \bibinfo
  {author} {\bibfnamefont {P.~J.}\ \bibnamefont {Steinhardt}},\ }\href
  {\doibase 10.1103/PhysRevLett.93.201301} {\bibfield  {journal} {\bibinfo
  {journal} {Phys.Rev.Lett.}\ }\textbf {\bibinfo {volume} {93}},\ \bibinfo
  {pages} {201301} (\bibinfo {year} {2004})},\ \Eprint
  {http://arxiv.org/abs/hep-ph/0403019} {arXiv:hep-ph/0403019 [hep-ph]}
  \BibitemShut {NoStop}%
\bibitem [{\citenamefont {Yanagida}(1979)}]{Yanagid1979}%
  \BibitemOpen
  \bibfield  {author} {\bibinfo {author} {\bibfnamefont {T.}~\bibnamefont
  {Yanagida}},\ }in\ \href@noop {} {\emph {\bibinfo {booktitle} {Proceedings of
  the Workshop on the Unified Theory and the Baryon Number in the Universe,
  Tsukuba, Japan, 1979}}},\ \bibinfo {editor} {edited by\ \bibinfo {editor}
  {\bibfnamefont {O.}~\bibnamefont {Sawada}}\ and\ \bibinfo {editor}
  {\bibfnamefont {A.}~\bibnamefont {Sugamoto}}}\ (\bibinfo  {publisher} {KEK},\
  \bibinfo {address} {Tsukuba},\ \bibinfo {year} {1979})\ p.~\bibinfo {pages}
  {95}\BibitemShut {NoStop}%
\bibitem [{\citenamefont {Yanagida}(1980)}]{Yanagida:1980xy}%
  \BibitemOpen
  \bibfield  {author} {\bibinfo {author} {\bibfnamefont {T.}~\bibnamefont
  {Yanagida}},\ }\href {\doibase 10.1143/PTP.64.1103} {\bibfield  {journal}
  {\bibinfo  {journal} {Prog. Theor. Phys.}\ }\textbf {\bibinfo {volume}
  {64}},\ \bibinfo {pages} {1103} (\bibinfo {year} {1980})}\BibitemShut
  {NoStop}%
\bibitem [{\citenamefont {Gell-Mann}\ \emph {et~al.}(1979)\citenamefont
  {Gell-Mann}, \citenamefont {Ramond},\ and\ \citenamefont
  {Slansky}}]{Gell-Mann1979}%
  \BibitemOpen
  \bibfield  {author} {\bibinfo {author} {\bibfnamefont {M.}~\bibnamefont
  {Gell-Mann}}, \bibinfo {author} {\bibfnamefont {P.}~\bibnamefont {Ramond}}, \
  and\ \bibinfo {author} {\bibfnamefont {R.}~\bibnamefont {Slansky}},\ }in\
  \href@noop {} {\emph {\bibinfo {booktitle} {Supergravity}}},\ \bibinfo
  {editor} {edited by\ \bibinfo {editor} {\bibfnamefont {D.}~\bibnamefont
  {Freedman}}\ and\ \bibinfo {editor} {\bibfnamefont {P.~V.}\ \bibnamefont
  {Nieuwenhuizen}}}\ (\bibinfo  {publisher} {North-Holland},\ \bibinfo
  {address} {Amsterdam},\ \bibinfo {year} {1979})\ pp.\ \bibinfo {pages}
  {315--321}\BibitemShut {NoStop}%
\bibitem [{\citenamefont {Fukugita}\ and\ \citenamefont
  {Yanagida}(1986)}]{Fukugita:1986hr}%
  \BibitemOpen
  \bibfield  {author} {\bibinfo {author} {\bibfnamefont {M.}~\bibnamefont
  {Fukugita}}\ and\ \bibinfo {author} {\bibfnamefont {T.}~\bibnamefont
  {Yanagida}},\ }\href {\doibase 10.1016/0370-2693(86)91126-3} {\bibfield
  {journal} {\bibinfo  {journal} {Phys. Lett.}\ }\textbf {\bibinfo {volume}
  {B174}},\ \bibinfo {pages} {45} (\bibinfo {year} {1986})}\BibitemShut
  {NoStop}%
\bibitem [{\citenamefont {Giudice}\ \emph {et~al.}(2004)\citenamefont
  {Giudice}, \citenamefont {Notari}, \citenamefont {Raidal}, \citenamefont
  {Riotto},\ and\ \citenamefont {Strumia}}]{Giudice:2003jh}%
  \BibitemOpen
  \bibfield  {author} {\bibinfo {author} {\bibfnamefont {G.}~\bibnamefont
  {Giudice}}, \bibinfo {author} {\bibfnamefont {A.}~\bibnamefont {Notari}},
  \bibinfo {author} {\bibfnamefont {M.}~\bibnamefont {Raidal}}, \bibinfo
  {author} {\bibfnamefont {A.}~\bibnamefont {Riotto}}, \ and\ \bibinfo {author}
  {\bibfnamefont {A.}~\bibnamefont {Strumia}},\ }\href {\doibase
  10.1016/j.nuclphysb.2004.02.019} {\bibfield  {journal} {\bibinfo  {journal}
  {Nucl. Phys.}\ }\textbf {\bibinfo {volume} {B685}},\ \bibinfo {pages} {89}
  (\bibinfo {year} {2004})},\ \Eprint {http://arxiv.org/abs/hep-ph/0310123}
  {arXiv:hep-ph/0310123 [hep-ph]} \BibitemShut {NoStop}%
\bibitem [{\citenamefont {Kearney}\ \emph {et~al.}(2015)\citenamefont
  {Kearney}, \citenamefont {Yoo},\ and\ \citenamefont
  {Zurek}}]{Kearney:2015vba}%
  \BibitemOpen
  \bibfield  {author} {\bibinfo {author} {\bibfnamefont {J.}~\bibnamefont
  {Kearney}}, \bibinfo {author} {\bibfnamefont {H.}~\bibnamefont {Yoo}}, \ and\
  \bibinfo {author} {\bibfnamefont {K.~M.}\ \bibnamefont {Zurek}},\ }\href@noop
  {} {\  (\bibinfo {year} {2015})},\ \Eprint {http://arxiv.org/abs/1503.05193}
  {arXiv:1503.05193 [hep-th]} \BibitemShut {NoStop}%
\bibitem [{\citenamefont {Cohen}\ \emph {et~al.}(1991)\citenamefont {Cohen},
  \citenamefont {Kaplan},\ and\ \citenamefont {Nelson}}]{Cohen:1990it}%
  \BibitemOpen
  \bibfield  {author} {\bibinfo {author} {\bibfnamefont {A.~G.}\ \bibnamefont
  {Cohen}}, \bibinfo {author} {\bibfnamefont {D.~B.}\ \bibnamefont {Kaplan}}, \
  and\ \bibinfo {author} {\bibfnamefont {A.~E.}\ \bibnamefont {Nelson}},\
  }\href {\doibase 10.1016/0550-3213(91)90395-E} {\bibfield  {journal}
  {\bibinfo  {journal} {Nucl. Phys.}\ }\textbf {\bibinfo {volume} {B349}},\
  \bibinfo {pages} {727} (\bibinfo {year} {1991})}\BibitemShut {NoStop}%
\bibitem [{\citenamefont {Dolgov}\ and\ \citenamefont
  {Freese}(1995)}]{Dolgov:1994zq}%
  \BibitemOpen
  \bibfield  {author} {\bibinfo {author} {\bibfnamefont {A.}~\bibnamefont
  {Dolgov}}\ and\ \bibinfo {author} {\bibfnamefont {K.}~\bibnamefont
  {Freese}},\ }\href {\doibase 10.1103/PhysRevD.51.2693} {\bibfield  {journal}
  {\bibinfo  {journal} {Phys.Rev.}\ }\textbf {\bibinfo {volume} {D51}},\
  \bibinfo {pages} {2693} (\bibinfo {year} {1995})},\ \Eprint
  {http://arxiv.org/abs/hep-ph/9410346} {arXiv:hep-ph/9410346 [hep-ph]}
  \BibitemShut {NoStop}%
\bibitem [{\citenamefont {Dolgov}\ \emph {et~al.}(1997)\citenamefont {Dolgov},
  \citenamefont {Freese}, \citenamefont {Rangarajan},\ and\ \citenamefont
  {Srednicki}}]{Dolgov:1996qq}%
  \BibitemOpen
  \bibfield  {author} {\bibinfo {author} {\bibfnamefont {A.}~\bibnamefont
  {Dolgov}}, \bibinfo {author} {\bibfnamefont {K.}~\bibnamefont {Freese}},
  \bibinfo {author} {\bibfnamefont {R.}~\bibnamefont {Rangarajan}}, \ and\
  \bibinfo {author} {\bibfnamefont {M.}~\bibnamefont {Srednicki}},\ }\href
  {\doibase 10.1103/PhysRevD.56.6155} {\bibfield  {journal} {\bibinfo
  {journal} {Phys.Rev.}\ }\textbf {\bibinfo {volume} {D56}},\ \bibinfo {pages}
  {6155} (\bibinfo {year} {1997})},\ \Eprint
  {http://arxiv.org/abs/hep-ph/9610405} {arXiv:hep-ph/9610405 [hep-ph]}
  \BibitemShut {NoStop}%
\bibitem [{\citenamefont {Dolgov}(1997)}]{Dolgov:1997qr}%
  \BibitemOpen
  \bibfield  {author} {\bibinfo {author} {\bibfnamefont {A.}~\bibnamefont
  {Dolgov}},\ }\href@noop {} {\  (\bibinfo {year} {1997})},\ \Eprint
  {http://arxiv.org/abs/hep-ph/9707419} {arXiv:hep-ph/9707419 [hep-ph]}
  \BibitemShut {NoStop}%
\bibitem [{\citenamefont {Dreiner}\ \emph {et~al.}(2010)\citenamefont
  {Dreiner}, \citenamefont {Haber},\ and\ \citenamefont
  {Martin}}]{Dreiner:2008tw}%
  \BibitemOpen
  \bibfield  {author} {\bibinfo {author} {\bibfnamefont {H.~K.}\ \bibnamefont
  {Dreiner}}, \bibinfo {author} {\bibfnamefont {H.~E.}\ \bibnamefont {Haber}},
  \ and\ \bibinfo {author} {\bibfnamefont {S.~P.}\ \bibnamefont {Martin}},\
  }\href {\doibase 10.1016/j.physrep.2010.05.002} {\bibfield  {journal}
  {\bibinfo  {journal} {Phys. Rept.}\ }\textbf {\bibinfo {volume} {494}},\
  \bibinfo {pages} {1} (\bibinfo {year} {2010})},\ \Eprint
  {http://arxiv.org/abs/0812.1594} {arXiv:0812.1594 [hep-ph]} \BibitemShut
  {NoStop}%
\bibitem [{\citenamefont {Nowakowski}\ and\ \citenamefont
  {Pilaftsis}(1993)}]{Nowakowski:1993iu}%
  \BibitemOpen
  \bibfield  {author} {\bibinfo {author} {\bibfnamefont {M.}~\bibnamefont
  {Nowakowski}}\ and\ \bibinfo {author} {\bibfnamefont {A.}~\bibnamefont
  {Pilaftsis}},\ }\href {\doibase 10.1007/BF01650437} {\bibfield  {journal}
  {\bibinfo  {journal} {Z. Phys.}\ }\textbf {\bibinfo {volume} {C60}},\
  \bibinfo {pages} {121} (\bibinfo {year} {1993})},\ \Eprint
  {http://arxiv.org/abs/hep-ph/9305321} {arXiv:hep-ph/9305321 [hep-ph]}
  \BibitemShut {NoStop}%
\bibitem [{\citenamefont {Cannoni}(2014)}]{Cannoni:2013zya}%
  \BibitemOpen
  \bibfield  {author} {\bibinfo {author} {\bibfnamefont {M.}~\bibnamefont
  {Cannoni}},\ }\href {\doibase 10.1103/PhysRevD.89.103533} {\bibfield
  {journal} {\bibinfo  {journal} {Phys. Rev.}\ }\textbf {\bibinfo {volume}
  {D89}},\ \bibinfo {pages} {103533} (\bibinfo {year} {2014})},\ \Eprint
  {http://arxiv.org/abs/1311.4494} {arXiv:1311.4494 [astro-ph.CO]} \BibitemShut
  {NoStop}%
\end{thebibliography}%

\end{document}